\pdfoutput=1
\documentclass[sort&compress,UKenglish,twocolumn,9pt]{article}

\usepackage[T3,T1]{fontenc}
\usepackage[utf8]{inputenc}
\usepackage{graphicx}
\usepackage{caption}
\usepackage{subcaption}
\usepackage{adjustbox}
\usepackage{siunitx}
\usepackage{bm}
\usepackage{amsmath}
\usepackage{amssymb}
\usepackage{fixmath}
\usepackage{mathtools}
\usepackage{nicefrac}
\usepackage{upgreek}
\usepackage{chemformula}

\DeclareSymbolFont{tipa}{T3}{cmr}{m}{n}
\DeclareMathAccent{\invbreve}{\mathalpha}{tipa}{16}

\usepackage[big,online]{dgruyter}
\makeatletter\AtBeginDocument{\sbox\dg@wordmark{}}\makeatother

\usepackage{glossaries-extra}
\makeglossaries
\setabbreviationstyle[acronym]{long-short}

\usepackage[%
    backend=biber,
    style=numeric,
    natbib=true,
    hyperref=true,
    backref=false,
    isbn=false,
    url=false,
    doi=true,
    urldate=long,
    maxnames=3,
    minnames=1,
    maxbibnames=5,
    minbibnames=3,
    maxcitenames=2,
    mincitenames=1,
    sorting=nty
]{biblatex}
\bibliography{bib/sources.bib}

\usepackage{hyperref}

\providecommand*{\figrefname}{Fig.~}
\newcommand*{\figref}[1]{%
  \hyperref[#1]{\figrefname{}}\ref{#1}%
}
\providecommand*{\tabrefname}{Tab.~}
\newcommand*{\tabref}[1]{%
  \hyperref[#1]{\tabrefname{}}\ref{#1}%
}
\providecommand*{\secrefname}{Section~}
\newcommand*{\secref}[1]{%
  \hyperref[#1]{\secrefname{}}\ref{#1}%
}

\newtheorem{lemma}{Lemma}[section]

\usepackage{tikz}

\usepackage{pgfplots}
\usetikzlibrary{plotmarks,math,positioning,shapes,arrows,backgrounds,circuits.logic.IEC,circuits.ee.IEC,decorations.text,decorations.pathmorphing,decorations.pathreplacing,decorations.markings,patterns,shapes.misc,shapes.geometric,calc,fit,spy,matrix,tikzmark,arrows.meta,calligraphy,bending}

\newcommand\varrightarrow[1][1.4em]{\tikz[baseline=-0.5ex, shorten <=2pt, shorten >=2pt] \draw[-latex'] (0,0) -- (#1,0);}
\newcommand\varmapsto[1][1.4em]{\tikz[baseline=-0.5ex, shorten <=2pt, shorten >=2pt] \draw[|-latex'] (0,0) -- (#1,0);}

\definecolor{mplblue}{RGB}{31, 119, 180}
\definecolor{mplorange}{RGB}{255, 127, 14}
\definecolor{mplgreen}{RGB}{44, 160, 44}
\definecolor{mplred}{RGB}{214, 39, 40}
\definecolor{mplpurple}{RGB}{148, 103, 189}
\definecolor{mplbrown}{RGB}{140, 86, 75}
\definecolor{mplpink}{RGB}{227, 119, 194}
\definecolor{mplgray}{RGB}{127, 127, 127}
\definecolor{mplolive}{RGB}{188, 189, 34}
\definecolor{mplcyan}{RGB}{23, 190, 207}
\definecolor{mplmagenta}{HTML}{FF00FF}

\DeclareRobustCommand{\plotMagentaLine}{\raisebox{2pt}{\tikz{\draw[mplmagenta,solid,line width=0.9pt](0,0) -- (5mm,0);}}}

\definecolor{bmimBlue}{cmyk}{1, .49, .0, .6}

\newcommand{\tikznode}[2]{\relax
    \ifmmode%
    \tikz[remember picture,baseline=(#1.base),inner sep=0pt] \node (#1) {$#2$};
    \else
    \tikz[remember picture,baseline=(#1.base),inner sep=0pt] \node (#1) {#2};%
    \fi
}

\tikzstyle{block} = [draw, fill=white, rectangle, minimum height=3em, minimum width=5.5cm]
\tikzstyle{rblock} = [draw, fill=white, rectangle, minimum height=3em, minimum width=5.5cm, text width=5.5cm]
\tikzstyle{nlblock} = [draw, postaction={draw,line width=0.25mm,white}, line width=0.5mm, black, fill=white, rectangle, minimum height=3em, minimum width=5em]
\tikzstyle{sum} = [draw, circle]
\tikzstyle{branch} = [circle,inner sep=0pt,minimum size=1mm,fill=black,draw=black, anchor=mid]
\tikzstyle{nvbranch} = [circle,inner sep=0pt,minimum size=1mm,fill=white,draw=white, fill opacity=0, draw opacity=0]
\tikzstyle{vecBranch} = [circle,inner sep=0pt,minimum size=2mm,fill=black,draw=black]
\tikzstyle{input} = [coordinate]
\tikzstyle{output} = [coordinate]
\tikzstyle{coord} = [coordinate]
\tikzset{
    box around/.style={
        draw,rounded corners,
        inner sep=2pt,outer sep=0pt,
        node contents={},fit=#1
    },
    dualblock/.style={
        shape = rectangle,
        rectangle split,
        rectangle split parts=2,
        draw=black,
        minimum height=2em,
        text width=7.5cm,
        inner sep=2pt,
        text centered,
    },
    boxtitle/.style={
        fill=bmimBlue,
        text=white,
        font=\scriptsize\bfseries
    },
}

\makeatletter
\newcommand*{\defeq}{\mathrel{\rlap{%
                    \raisebox{0.3ex}{$\m@th\cdot$}}%
                    \raisebox{-0.3ex}{$\m@th\cdot$}}%
                    =}
\newcommand*{\defeqq}{=\mathrel{\rlap{%
                      \raisebox{0.3ex}{$\m@th\cdot$}}%
                     \raisebox{-0.3ex}{$\m@th\cdot$}}}
\makeatother

\newcommand\ttext[1]{\mathchoice%
  {\text{#1}}%
  {\text{#1}}%
  {\thickmuskip=1mu\medmuskip=1mu\text{#1}}%
  {\thickmuskip=1mu\medmuskip=1mu\text{#1}}}

\newcommand{\deri}[2][non]{\ensuremath{ \tfrac{\mathrm{d}\, #1}{\mathrm{d}\, #2} }}
\newcommand{\pderi}[1][non]{\ensuremath{\ifthenelse{\equal{#1}{non}}{\partial}}{\partial_{#1}}}
\newcommand{\dd}[1]{{\textrm d}#1}

\renewcommand{\Re}{\ensuremath{\mathbb{R}}}
\newcommand{\Co}{\ensuremath{\mathbb{C}}}
\newcommand{\N}{\ensuremath{\mathbb{N}}}
\newcommand{\trans}{\ensuremath{\intercal}}

\newcommand{\spec}[1][non]{\ensuremath{\ifthenelse{\equal{#1}{non}}{\sigma}{\sigma_{\ttext{#1}}}}}

\newcommand{\Leb}[1][non]{\ensuremath{\ifthenelse{\equal{#1}{non}}{L}{L^{\ttext{#1}}}}}
\newcommand{\Hil}[1][non]{\ensuremath{\ifthenelse{\equal{#1}{non}}{H}{H^{\ttext{#1}}}}}
\newcommand{\Cont}[1][non]{\ensuremath{\ifthenelse{\equal{#1}{non}}{C}{C^{\ttext{#1}}}}}

\newcommand{\X}[1][non]{\ensuremath{\ifthenelse{\equal{#1}{non}}{X}{X_{\ttext{#1}}}}}
\newcommand{\Xn}[1][non]{\ensuremath{\ifthenelse{\equal{#1}{non}}{X^{0}}{X^{0}_{\ttext{#1}}}}}
\newcommand{\Yn}[1][non]{\ensuremath{\ifthenelse{\equal{#1}{non}}{Y^{0}}{Y^{0}_{\ttext{#1}}}}}
\newcommand{\Xo}[1][non]{\ensuremath{\ifthenelse{\equal{#1}{non}}{X^{\perp}}{X^{\perp}_{\ttext{#1}}}}}
\newcommand{\Yo}[1][non]{\ensuremath{\ifthenelse{\equal{#1}{non}}{Y^{\perp}}{Y^{\perp}_{\ttext{#1}}}}}

\newcommand{\bmf}[1][non]{\ensuremath{\ifthenelse{\equal{#1}{non}}{\bm{\mathrm{f}}}{\bm{\mathrm{f}}_{\ttext{#1}}}}}
\newcommand{\stf}[1][non]{\ensuremath{\ifthenelse{\equal{#1}{non}}{\mathrm{f}}{\mathrm{f}_{\ttext{#1}}}}}
\newcommand{\stg}[1][non]{\ensuremath{\ifthenelse{\equal{#1}{non}}{\mathrm{g}}{\mathrm{g}_{\ttext{#1}}}}}
\newcommand{\sth}[1][non]{\ensuremath{\ifthenelse{\equal{#1}{non}}{\mathrm{h}}{\mathrm{h}_{\ttext{#1}}}}}
\newcommand{\bmA}[1][non]{\ensuremath{\ifthenelse{\equal{#1}{non}}{\bm{\mathrm{A}}}{\bm{\mathrm{A}}_{\ttext{#1}}}}}
\newcommand{\hbmA}[1][non]{\ensuremath{\ifthenelse{\equal{#1}{non}}{\hat{\bm{\mathrm{A}}}}{\hat{\bm{\mathrm{A}}}_{\ttext{#1}}}}}
\newcommand{\bmB}[1][non]{\ensuremath{\ifthenelse{\equal{#1}{non}}{\bm{\mathrm{B}}}{\bm{\mathrm{B}}_{\ttext{#1}}}}}
\newcommand{\hbmB}[1][non]{\ensuremath{\ifthenelse{\equal{#1}{non}}{\hat{\bm{\mathrm{B}}}}{\hat{\bm{\mathrm{B}}}_{\ttext{#1}}}}}
\newcommand{\B}[1][non]{\ensuremath{\ifthenelse{\equal{#1}{non}}{\mathrm{B}}{\mathrm{B}_{\ttext{#1}}}}}

\newcommand{\opI}{\ensuremath{\mathcal{I}}}

\newcommand{\opA}[1][non]{\ensuremath{\ifthenelse{\equal{#1}{non}}{\mathcal{A}}{\mathcal{A}_{\ttext{#1}}}}}
\newcommand{\opB}[1][non]{\ensuremath{\ifthenelse{\equal{#1}{non}}{\mathcal{B}}{\mathcal{B}_{\ttext{#1}}}}}
\newcommand{\opC}{\ensuremath{\mathcal{C}}}
\newcommand{\opD}[1][non]{\ensuremath{\ifthenelse{\equal{#1}{non}}{\mathcal{D}}{\mathcal{D}_{\ttext{#1}}}}}
\newcommand{\opP}[1][non]{\ensuremath{\ifthenelse{\equal{#1}{non}}{\mathcal{P}}{\mathcal{P}_{\ttext{#1}}}}}
\newcommand{\opK}[1][non]{\ensuremath{\ifthenelse{\equal{#1}{non}}{\mathcal{K}}{\mathcal{K}_{\ttext{#1}}}}}
\newcommand{\opR}[1][non]{\ensuremath{\ifthenelse{\equal{#1}{non}}{\mathcal{R}}{\mathcal{R}_{\ttext{#1}}}}}

\newcommand{\bmI}[1][non]{\ensuremath{\ifthenelse{\equal{#1}{non}}{\bm{\mathrm{I}}}{\bm{\mathrm{I}}_{\ttext{#1}}}}}
\newcommand{\bmL}[1][non]{\ensuremath{\ifthenelse{\equal{#1}{non}}{\bm{\mathrm{L}}}{\bm{\mathrm{L}}_{\ttext{#1}}}}}
\newcommand{\bmR}[1][non]{\ensuremath{\ifthenelse{\equal{#1}{non}}{\bm{\mathrm{R}}}{\bm{\mathrm{R}}_{\ttext{#1}}}}}

\newcommand{\bmu}[1][non]{\ensuremath{\ifthenelse{\equal{#1}{non}}{\bm{u}}{\bm{u}_{\ttext{#1}}}}}
\newcommand{\bbmu}[1][non]{\ensuremath{\ifthenelse{\equal{#1}{non}}{\bm{\bar{u}}}{\bm{\bar{u}}_{\ttext{#1}}}}}
\newcommand{\bmInputs}[1][non]{\ensuremath{\ifthenelse{\equal{#1}{non}}{\bm{\upsilon}}{\bm{\upsilon}_{\ttext{#1}}}}}

\newcommand{\bmV}[1][non]{\ensuremath{\ifthenelse{\equal{#1}{non}}{\bm{\mathrm{V}}}{\bm{\mathrm{V}}_{\ttext{#1}}}}}
\newcommand{\bmVo}[1][non]{\ensuremath{\ifthenelse{\equal{#1}{non}}{\bm{\mathrm{V}}^{\perp}}{\bm{\mathrm{V}}^{\perp}_{\ttext{#1}}}}}
\newcommand{\bmW}[1][non]{\ensuremath{\ifthenelse{\equal{#1}{non}}{\bm{\mathrm{W}}}{\bm{\mathrm{W}}_{\ttext{#1}}}}}
\newcommand{\bmO}[1][non]{\ensuremath{\ifthenelse{\equal{#1}{non}}{\bm{\mathrm{O}}}{\bm{\mathrm{O}}_{\ttext{#1}}}}}
\newcommand{\bmG}[1][non]{\ensuremath{\ifthenelse{\equal{#1}{non}}{\bm{\mathrm{G}}}{\bm{\mathrm{G}}_{\ttext{#1}}}}}

\newcommand{\uOmega}[1][non]{\ensuremath{\ifthenelse{\equal{#1}{non}}{\upOmega}{\upOmega_{\ttext{#1}}}}}
\newcommand{\odeFlow}[1][non]{\ensuremath{\ifthenelse{\equal{#1}{non}}{\Upphi}{\Upphi_{\ttext{#1}}}}}
\newcommand{\odesFlow}[1][non]{\ensuremath{\ifthenelse{\equal{#1}{non}}{\bm{\Upphi}}{\bm{\Upphi}_{\ttext{#1}}}}}
\newcommand{\hOdesFlow}[1][non]{\ensuremath{\ifthenelse{\equal{#1}{non}}{\bm{\hat{\Upphi}}}{\bm{\hat{\Upphi}}_{\ttext{#1}}}}}
\newcommand{\cOdesFlow}[1][non]{\ensuremath{\ifthenelse{\equal{#1}{non}}{\bm{\check{\Upphi}}}{\bm{\check{\Upphi}}_{\ttext{#1}}}}}
\newcommand{\spatDom}{\ensuremath{\uOmega[$z$]}}

\renewcommand{\ker}[1][non]{\ensuremath{\ifthenelse{\equal{#1}{non}}{\mathrm{kern}}{\mathrm{kern}_{\ttext{#1}}}}}
\newcommand{\im}[1][non]{\ensuremath{\ifthenelse{\equal{#1}{non}}{\mathrm{im}}{\mathrm{im}_{\ttext{#1}}}}}
\newcommand{\rank}[1][non]{\ensuremath{\ifthenelse{\equal{#1}{non}}{\mathrm{rank}}{\mathrm{rank}_{\ttext{#1}}}}}

\newcommand{\bbmx}[1][non]{\ensuremath{\ifthenelse{\equal{#1}{non}}{\bm{\bar{x}}}{\bm{\bar{x}}_{\ttext{#1}}}}}
\newcommand{\bmx}[1][non]{\ensuremath{\ifthenelse{\equal{#1}{non}}{\bm{x}}{\bm{x}_{\ttext{#1}}}}}
\newcommand{\del}[1][non]{\ensuremath{\ifthenelse{\equal{#1}{non}}{\delta}{\delta_{\ttext{#1}}}}}
\newcommand{\bmdel}[1][non]{\ensuremath{\ifthenelse{\equal{#1}{non}}{\bm{\delta}}{\bm{\delta}_{\ttext{#1}}}}}
\newcommand{\bdel}[1][non]{\ensuremath{\ifthenelse{\equal{#1}{non}}{\bar{\delta}}{\bar{\delta}_{\ttext{#1}}}}}
\newcommand{\delo}[1][non]{\ensuremath{\ifthenelse{\equal{#1}{non}}{\delta^{\perp}}{\delta^{\perp}_{\ttext{#1}}}}}
\newcommand{\bdelo}[1][non]{\ensuremath{\ifthenelse{\equal{#1}{non}}{\bar{\delta}^{\perp}}{\bar{\delta}^{\perp}_{\ttext{#1}}}}}
\newcommand{\bmdelo}[1][non]{\ensuremath{\ifthenelse{\equal{#1}{non}}{\bm{\delta}^{\perp}}{\bm{\delta}^{\perp}_{\ttext{#1}}}}}
\newcommand{\bbmdelo}[1][non]{\ensuremath{\ifthenelse{\equal{#1}{non}}{\bm{\bar{\delta}}^{\perp}}{\bm{\bar{\delta}}^{\perp}_{\ttext{#1}}}}}
\newcommand{\deln}[1][non]{\ensuremath{\ifthenelse{\equal{#1}{non}}{\delta^{0}}{\delta^{0}_{\ttext{#1}}}}}
\newcommand{\bdeln}[1][non]{\ensuremath{\ifthenelse{\equal{#1}{non}}{\bar{\delta}^{0}}{\bar{\delta}^{0}_{\ttext{#1}}}}}
\newcommand{\bbmdeln}[1][non]{\ensuremath{\ifthenelse{\equal{#1}{non}}{\bm{\bar{\delta}}^{0}}{\bm{\bar{\delta}}^{0}_{\ttext{#1}}}}}
\newcommand{\bmdelta}[1][non]{\ensuremath{\ifthenelse{\equal{#1}{non}}{\bm{\delta}}{\bm{\delta}_{\ttext{#1}}}}}
\newcommand{\bbmdelta}[1][non]{\ensuremath{\ifthenelse{\equal{#1}{non}}{\bar{\bm{\delta}}}{\bar{\bm{\delta}}_{\ttext{#1}}}}}
\newcommand{\Zeta}[1][non]{\ensuremath{\ifthenelse{\equal{#1}{non}}{\zeta}{\zeta_{\ttext{#1}}}}}
\newcommand{\bmZeta}[1][non]{\ensuremath{\ifthenelse{\equal{#1}{non}}{\bm{\zeta}}{\bm{\zeta}_{\ttext{#1}}}}}
\newcommand{\dtbmZeta}[1][non]{\ensuremath{\ifthenelse{\equal{#1}{non}}{\dot{\bm{\zeta}}}{\dot{\bm{\zeta}}_{\ttext{#1}}}}}
\newcommand{\Eta}[1][non]{\ensuremath{\ifthenelse{\equal{#1}{non}}{\eta}{\eta_{\ttext{#1}}}}}
\newcommand{\bmEta}[1][non]{\ensuremath{\ifthenelse{\equal{#1}{non}}{\bm{\eta}}{\bm{\eta}_{\ttext{#1}}}}}
\newcommand{\Nu}[1][non]{\ensuremath{\ifthenelse{\equal{#1}{non}}{\nu}{\nu_{\ttext{#1}}}}}
\newcommand{\bmNu}[1][non]{\ensuremath{\ifthenelse{\equal{#1}{non}}{\bm{\nu}}{\bm{\nu}_{\ttext{#1}}}}}
\newcommand{\lam}[1][non]{\ensuremath{\ifthenelse{\equal{#1}{non}}{\lambda}{\lambda_{\ttext{#1}}}}}
\newcommand{\bmlam}[1][non]{\ensuremath{\ifthenelse{\equal{#1}{non}}{\bm{\lambda}}{\bm{\lambda}_{\ttext{#1}}}}}
\newcommand{\fT}[1][non]{\ensuremath{\ifthenelse{\equal{#1}{non}}{\mathbf{T}}{\mathrm{T}_{\mathrm{#1}}}}}

\newcommand{\bx}[1][non]{\ensuremath{\ifthenelse{\equal{#1}{non}}{\bar{x}}{\bar{x}_{\ttext{#1}}}}}
\newcommand{\bg}[1][non]{\ensuremath{\ifthenelse{\equal{#1}{non}}{\bar{g}}{\bar{g}_{\ttext{#1}}}}}

\newcommand{\tsA}{\ensuremath{t_{\ttext{a}}}}
\newcommand{\tsB}{\ensuremath{t_{\ttext{b}}}}
\newcommand{\tS}{\ensuremath{t_{0}}}
\newcommand{\tM}{\ensuremath{t_{\ttext{m}}}}

\renewcommand{\ker}[1][non]{\ensuremath{\ifthenelse{\equal{#1}{non}}{\mathrm{kern}}{\mathrm{kern}_{\ttext{#1}}}}}
\newcommand{\diag}[1][non]{\ensuremath{\ifthenelse{\equal{#1}{non}}{\mathrm{diag}}{\mathrm{diag}_{\ttext{#1}}}}}
\newcommand{\spa}[1][non]{\ensuremath{\ifthenelse{\equal{#1}{non}}{\mathrm{span}}{\mathrm{span}_{\ttext{#1}}}}}

\newcommand{\Del}{\ensuremath{^{-}}}
\newcommand{\Pred}{\ensuremath{^{+}}}

\newcommand{\uPhi}[1][non]{\ensuremath{\ifthenelse{\equal{#1}{non}}{\mathrm{\Phi}}{\mathrm{\Phi}_{\ttext{#1}}}}}

\newcommand{\vel}[1][non]{\ensuremath{\ifthenelse{\equal{#1}{non}}{\lambda}{\lambda_{\ttext{#1}}}}}
\newcommand{\dvel}[1][non]{\ensuremath{\ifthenelse{\equal{#1}{non}}{\dot{\lambda}}{\dot{\lambda}_{\ttext{#1}}}}}

\newcommand{\bmw}[1][non]{\ensuremath{\ifthenelse{\equal{#1}{non}}{\bm{w}}{\bm{w}_{\ttext{#1}}}}}

\newcommand{\ctrlIn}[1][non]{\ensuremath{\ifthenelse{\equal{#1}{non}}{u}{u_{\ttext{#1}}}}}
\newcommand{\distIn}[1][non]{\ensuremath{\ifthenelse{\equal{#1}{non}}{d}{d_{\ttext{#1}}}}}
\newcommand{\bdistIn}[1][non]{\ensuremath{\ifthenelse{\equal{#1}{non}}{\bm{d}}{\bm{d}_{\ttext{#1}}}}}
\newcommand{\dtbdistIn}[1][non]{\ensuremath{\ifthenelse{\equal{#1}{non}}{\dot{\bm{d}}}{\dot{\bm{d}}_{\ttext{#1}}}}}
\newcommand{\dtbbdistIn}[1][non]{\ensuremath{\ifthenelse{\equal{#1}{non}}{\dot{\bar{\bm{d}}}}{\dot{\bar{\bm{d}}}_{\ttext{#1}}}}}
\newcommand{\ctrlOut}[1][non]{\ensuremath{\ifthenelse{\equal{#1}{non}}{y}{y_{\ttext{#1}}}}}
\newcommand{\dtCtrlOut}[1][non]{\ensuremath{\ifthenelse{\equal{#1}{non}}{\dot{y}}{\dot{y}_{\ttext{#1}}}}}
\newcommand{\dtbCtrlOut}[1][non]{\ensuremath{\ifthenelse{\equal{#1}{non}}{\dot{\bar{y}}}{\dot{\bar{y}}_{\ttext{#1}}}}}

\newcommand{\Kp}[1][non]{\ensuremath{\ifthenelse{\equal{#1}{non}}{K_{\ttext{P}}}{K_{\ttext{P, #1}}}}}
\newcommand{\Ki}[1][non]{\ensuremath{\ifthenelse{\equal{#1}{non}}{K_{\ttext{I}}}{K_{\ttext{I, #1}}}}}
\newcommand{\tKp}[1][non]{\ensuremath{\ifthenelse{\equal{#1}{non}}{\tilde{K}_{\ttext{P}}}{\tilde{K}_{\ttext{P, #1}}}}}
\newcommand{\tKi}[1][non]{\ensuremath{\ifthenelse{\equal{#1}{non}}{\tilde{K}_{\ttext{I}}}{\tilde{K}_{\ttext{I, #1}}}}}

\newacronym{pde}{PDE}{partial differential equation}
\newacronym{ode}{ODE}{ordinary differential equation}
\newacronym{dde}{DDE}{delay differential equation}
\newacronym{pdae}{PDAE}{partial differential-algebraic equation}
\newacronym{dfpde}{dFPDE}{delayed filter partial differential equation}
\newacronym{bvp}{BVP}{boundary value problem}
\newacronym{ivp}{IVP}{initial value problem}
\newacronym{scr}{SCR}{selective catalytic reduction}
\newacronym{fdm}{FDM}{finite-difference method}
\newacronym{bc}{BC}{boundary condition}
\newacronym{ic}{IC}{initial condition}
\newacronym{dae}{DAE}{differential-algebraic equation}

\makeglossaries

\begin{document}
  \selectlanguage{UKenglish}

  \author*[1]{Jens Wurm}
  \author[2]{Frank  Woittennek}
  \runningauthor{Wurm et al.}

  \affil[1]{Institute of Automation and Control Engineering, UMIT TIROL, 6060 Hall in
  Tirol, Austria, Email: jens.wurm@umit-tirol.at}
  \affil[2]{Institute of Automation and Control Engineering, UMIT TIROL, 6060 Hall in
  Tirol, Austria, Email: frank.woittennek@umit-tirol.at}

  \title{Control design for a class of nonlinear convection-reaction systems with storage effects and relative degree one}

  \runningtitle{Control design of convection-reaction models with storage effects}

  \abstract{
    This article addresses the control design for systems of several con\-vec\-tion-reaction equations distributively coupled to a reaction
    equation.
    In typical applications, input and output are associated exclusively with the convection-reaction subsystems.
    Along the common characteristics of the transport processes, the system can be reduced to a single delayed evolutionary \glsxtrshort{ode}.
    Based on the latter representation, an inversion-based feedforward control is proposed, whose properties depend on the coupling structure of the
    original system.
    While the relative degree of the output is zero in many cases, a relative degree of one is of particular interest for practical applications.
    To compensate for uncertainties and disturbances, the feedforward control is extended by a tracking controller with a predictor based on
    the transformed system to account for the resulting delays.
  }

  \keywords{hyperbolic system, input-output linearisation, distributed parameter systems, predictor control}

  \transabstract[ngerman]{
    Dieser Beitrag behandelt den Reglerentwurf für Systeme, die aus mehreren Konvektions-Reaktions-Gleichungen bestehen, welche verteilt mit einer
    Reaktionsgleichung gekoppelt sind.
    In typischen Anwendungen sind sowohl der Systemeingang als auch der Systemausgang ausschließlich den Konvektions-Reaktions-Teilsystemen
    zugeordnet.
    Entlang der gemeinsamen Charakteristiken der Transportprozesse lässt sich das System auf eine einzelne zeitverzögerte gewöhnliche
    Differentialgleichung reduzieren.
    Auf Basis dieser Darstellung wird eine inversionsbasierte Steuerung vorgeschlagen, deren Eigenschaften von der Kopplungsstruktur des
    ursprünglichen Systems abhängen.
    Während der relative Grad des Ausgangs in vielen Fällen null ist, ist für praktische Anwendungen ein relativer Grad von eins von besonderem Interesse.
    Zur Kompensation von Unbestimmtheiten und unbekannten Störungen wird die Steuerung um einen Folgeregler erweitert, der einen Prädiktor auf Basis des
    transformierten Systems beinhaltet, um die resultierenden Totzeiten zu berücksichtigen.
  }

  \transkeywords[ngerman]{hyperbolische Systeme, Eingangs-Ausgangs-Linearisierung, verteiltparametrische Systeme, Prädiktorregelung}

  \startpage{1}

\maketitle

\section{Introduction}
\label{sec:intro}

Systems that combine convection-reaction processes with a reaction or storage process coupled in a distributed manner arise in several technical
domains.
Typical examples include pipe models in cooling loops \cite{Bachler2017}, solar desalination plants \cite{Santos2011}, and catalysts for exhaust gas
aftertreatment \cite{Devarakonda2008,Yuan2015}.
In these systems, both boundary control and boundary measurement are associated with the variables of the transport processes, whereas the variable
associated with the storage process can neither be influenced nor measured directly.
The remaining inflows act as a priori known disturbances.

In \cite{Wurm2020}, the authors proposed a concept for a single transport process that completely decouples the transport process from the storage
process.
This concept was employed for an efficient feedforward design and was subsequently generalised in \cite{Wurm2023} to systems involving an
arbitrary number of transport processes, albeit under the assumption of uniform velocity.
In a further contribution \cite{Wurm2024}, the coupling structure of the convection-reaction system was examined in greater detail, demonstrating
that the properties of the input-output behaviour, and consequently those of its inverse, are governed by this coupling structure.
In contrast to these previous studies, which considered purely linear systems, the present work focuses on nonlinear reaction terms.
An alternative modelling approach is presented in \cite{BreschPietri2019} for linear hyperbolic systems with storage dynamics by transforming them
into a convection-diffusion equation.

The control design is achieved by rewriting the system dynamics along the characteristics of the transport processes, yielding a delayed
semi-explicit \gls{pdae} system composed of \glspl{ode} w.r.t. the spatial domain that describe the convection-reaction processes.
These spatial \glspl{ode} remain distributively coupled with an \gls{ode} w.r.t. time that describes the storage process.
By taking into account the solution of the spatial \glspl{ode}, the transport processes can be separated from the storage dynamics, yielding a single
delayed evolutionary \gls{ode}.
In \cite{Bachler2020}, this equation was termed \gls{dfpde} to reflect both the (hidden) time delays and an input-output behaviour reminiscent of an
infinite-dimensional low-pass filter.

The latter equation forms the starting point for the inversion-based control design.
The inverse model is again given by a delayed evolutionary \gls{ode} of the same type that accounts for the system in which the output assumes the
role of the input.
The properties reported in \cite{Wurm2024} regarding the coupling structure of the original convection-reaction system are extended here to the
nonlinear case.
Accordingly, the relative degree of the output for the delayed system is zero in many cases, but becomes one through an appropriate coupling
structure.
In this context, the \gls{ode} describes the internal dynamics.

The resulting calculation rule for the control input, derived from the solution of the \gls{ode}, is used to directly construct a tracking
controller that compensates for parameter uncertainties and unknown disturbances.
The resulting control law can be interpreted in terms of input-output linearisation \cite{Slotine1991} of the system.
Since the transport delays must be accounted for, predicted values are required within the controller.
To this end, a predictor-corrector-based control scheme is proposed that follows the structure of the Smith predictor \cite{Smith1957} and, unlike the
state-feedback approach in \cite{Fridman2014}, does not require an observer to reconstruct the entire state.

Related control approaches for transport systems without storage dynamics and with spatially distributed control input can be found, for instance,
in \cite{Alt2014}.
Alternatively, \cite{Zimmer2025} presents a backstepping-based controller design for linear, single-sided, fully actuated \glspl{pdae}.

The contribution is structured as follows: \secref{sec:problem} introduces the considered system class and its representation along the
characteristics.
\secref{sec:feedForward} presents the feedforward design, first for the linear case, including the solvability w.r.t. the input as well as the
well-posedness and stability of the internal dynamics, and then for the nonlinear case.
The tracking controller with the predictor is derived in \secref{sec:controller}, and \secref{sec:example} illustrates the results by an example
with two transport processes, focusing on the implementation.

\textit{Notation:}
The sets of real numbers and nonnegative real numbers are indicated by $\Re$ and $\Re^{+}$.
The first derivative of $x(t)$ w.r.t. time is denoted by $\dot{x}(t)$ and the first derivative of $x(z)$ w.r.t. the spatial variable by $x'(z)$.
For the partial derivative of $x(z,t)$ w.r.t. $z$ the notation $\pderi[z] x(z,t)$ is used as well as for the partial derivative of $x(z,t)$ w.r.t. $t$
$\pderi[t] x(z,t)$.
The shorthand $\Cont[$k$](\spatDom)$ represents the space of $k$ times continuously differentiable functions on domain $\spatDom$, where
$\Cont[$0$](\spatDom)$ is equipped with the supremum norm.
The diacritical mark $\bar{\square}$ denotes function arguments; for the spatially bundled states $\bar{\bmZeta} \defeq \bmZeta(\cdot, t) \in
\Cont[$1$](\spatDom, \Re^{n})$ and $\bar{\Eta} \defeq \Eta(\cdot, t) \in \Cont[$0$](\spatDom)$, the spatial dependence is absorbed into the
function space.
The Gâteaux derivative of an operator $f$ at the point $\bar{x}$ in the direction $\delta \bar{x}$ is defined as $\pderi[\bar{x}] f(\bar{x})
\delta \bar{x} = \deri[]{\epsilon} f(\bar{x} + \epsilon\, \delta \bar{x})\Big|_{\epsilon = 0}$, cf. \cite{Zeidler1986}.

\section{Problem statement}
\label{sec:problem}

The processes considered in this article can all be described by one-dimensional \glspl{pde}, which comprise several convection-reaction processes that are
coupled with a further equation modelling the storage process.
The convection-reaction processes are convection-dominated processes, meaning that slow processes, such as heat conduction or diffusion, can be neglected.
Furthermore, it is assumed that the transport processes have the same characteristic velocity and direction.
Thus, the system under consideration can be expressed in the general nonlinear form
\begin{subequations}\label{eq:model}
\begin{align}
    \pderi[t] \bmZeta(z,t) & = -\vel(t) \pderi[z] \bmZeta(z,t) + \bmf(\bmZeta(z,t), \Eta(z,t)) \label{eq:model:transport}\\
    \pderi[t] \Eta(z,t) & = \stg(\bmZeta(z,t), \Eta(z,t)) \label{eq:model:storage}
\end{align}
with the \glspl{ic}
\begin{align}
    \bmZeta(z,t_0) & = \bmZeta[0](z), &
    \Eta(z,t_0) & = \Eta[0](z),
\end{align}
\end{subequations}
which is defined on the spatial domain $z \in \spatDom = [0, L]$ with $L > 0$ and the time $t \in \Re^{+}_{t_0} = \{t \in \Re^{+} \colon t > t_0\}$.
Therein, the variables $(z,t)\varmapsto\bmZeta(z,t)\in\Re^{n}$ are associated with the convection processes while $(z,t)\varmapsto\Eta(z,t)\in\Re$
denotes the storage variable.
Here, $\bmZeta(z,t)$ denotes isotachic states, i.e., states sharing a common, positive and sufficiently smooth transport velocity $0 < \underline{\vel}
\le \vel(t) \le \overline{\vel}$, where $\underline{\vel}$ and $\overline{\vel}$ are the lower and upper bounds, respectively.
The process equations are coupled by the reaction terms $\bmf \colon \Re^{n+1} \varrightarrow \Re^{n}$ and $\stg \colon \Re^{n+1} \varrightarrow
\Re$.

By means of a controller, arbitrary trajectories of the output $\ctrlOut(t) \in \Re$ for \eqref{eq:model} are to be achieved.
A typical example is the transition between the two stationary regimes $(\ctrlOut[a], \ctrlOut[b])$ on the interval $(\tsA,\tsB)$.
The output to be transferred $\ctrlOut(t)\in\Re$ and the control input $\ctrlIn(t) \in \Re$ correspond to linear combinations of the transport process variables
$\bmZeta(z,t)$ at $z=L$ and $z=0$, respectively:\\[-3ex]
\begin{subequations}\label{eq:model:inout}
\begin{align}
    \ctrlOut(t) & = \bmW \bmZeta(L,t), &
    \bmW & \in \Re^{1\times n}, \\
    \ctrlIn(t) & = \bmVo \bmZeta(0,t), &
    \bmVo & \in \Re^{1\times n}.
    \label{eq:model:inout:ctrlIn}
\end{align}
The disturbance variables $\bdistIn(t) \in \Re^{n-1}$, which cannot be influenced but are known a priori, are given analogously by
\begin{align}
    \bdistIn(t) & = \bmV \bmZeta(0,t), &
    \bmV & \in \Re^{(n-1)\times n}.
    \label{eq:model:inout:distIn}
\end{align}
\end{subequations}
Moreover, $\bmVo$ and the rows of $\bmV$ are assumed to be orthonormal, i.e., the matrix $((\bmVo)^{\trans}, \bmV^{\trans})^{\trans} \in \Re^{n
\times n}$ is orthogonal.
This can always be achieved by a suitable choice of the disturbance variables and a scaling of the control input, and yields the decomposition
$\bmZeta(0,t) = (\bmVo)^{\trans} \ctrlIn(t) + \bmV^{\trans} \bdistIn(t)$.

\subsection{Method of characteristics}

In order to simplify the further steps, it is necessary to reformulate system \eqref{eq:model} by transforming the transport processes
\eqref{eq:model:transport} into \glspl{ode} w.r.t. the spatial domain.
To this end, the method of characteristics is employed, see, e.g., \cite{Courant1937, Folland1976}.
This method allows the original system to be reformulated in a way that offers a more convenient approach to its solution and, furthermore, a
geometric interpretation \cite{John1975}.
By applying the method of characteristics to \eqref{eq:model}, one obtains the characteristic differential equations in terms of $z$ and $t$ with
\begin{align*}
    \deri[z]{\xi} (\xi) & = 1, &
    \deri[t]{\xi} (\xi) & = \tfrac{1}{\vel(t(\xi))}.
\end{align*}
The integration of both equations for the interval $[s,L]$, $s\in[0,L]$ yields, on the one hand, $z(L) - z(s) = L - s$, i.e., $z = s$, and, on
the other hand, an equation for the space- and time-dependent delay time $(z,t) \varmapsto \tau(L - z,t)$
\begin{align}
    \int_{t - \tau(L - z, t)}^{t} \vel(\xi) \dd \xi & = L - z.
    \label{eq:charSys:delay}
\end{align}
For simplicity, $\varphi(z,t) \defeq t - \tau(L - z, t)$ is introduced.
This allows the delayed variables
\begin{align*}
    \bmw\Del(z, t) & = \bm{w}(z, \varphi(z,t)), &
    \bmw & = (\bmZeta^{\trans}, \Eta)^{\trans}
\end{align*}
to be defined.
The spatial and temporal derivatives of the delayed quantities are given by
\begin{subequations}\label{eq:charSys:delsys:deriv}
\begin{align}
    \pderi[t] \bmw\Del(z,t) & = \pderi[t] \varphi(z,t) \pderi[\varphi] \bmw(z,\varphi(z,t)), \\
    \pderi[z] \bmw\Del(z,t) & = \pderi[z] \bmw(z,\varphi(z,t)) \notag\\
                            & \quad + \pderi[z] \varphi(z,t) \pderi[\varphi] \bmw(z,\varphi(z,t)).
\end{align}
\end{subequations}
To evaluate these expressions, the spatial and temporal derivatives of $\varphi(z,t)$ must be known.
For this purpose, \eqref{eq:charSys:delay} must be differentiated w.r.t. $t$ and $z$ by means of the Leibniz rule for parametric integrals,
yielding
\begin{align}
    \pderi[t] \varphi(z,t)  & = \tfrac{\lam(t)}{\lam\Del(z,t)}, &
    \pderi[z] \varphi(z,t) & = \tfrac{1}{\lam\Del(z,t)}
    \label{eq:charSys:delsys:int}
\end{align}
with delayed transport velocity $\vel\Del(z,t) \defeq \vel(\varphi(z,t))$.
Using \eqref{eq:charSys:delsys:int}, \eqref{eq:charSys:delsys:deriv} simplifies to
\begin{align*}
    \pderi[t] \bmw\Del(z,t) & = \tfrac{\lam(t)}{\lam\Del(z,t)} \pderi[\varphi] \bmw(z,\varphi(z,t)), \\
    \pderi[z] \bmw\Del(z,t) & = \pderi[z] \bmw(z,\varphi(z,t)) + \tfrac{1}{\lam\Del(z,t)} \pderi[\varphi] \bmw(z,\varphi(z,t)),
\end{align*}
which, when substituted into \eqref{eq:model}, gives the delayed semi-explicit \gls{pdae} system
\begin{subequations}\label{eq:model:del}
\begin{align}
    \pderi[z] \bmZeta\Del(z,t) & = \hat{\bmf}(\bmZeta\Del(z,t), \Eta\Del(z,t), \vel\Del(z,t)) \label{eq:model:del:transport}\\
    \pderi[t] \Eta\Del(z,t) & = \hat{\stg}(\bmZeta\Del(z,t), \Eta\Del(z,t), \vel\Del(z,t), \vel(t)) \label{eq:model:del:storage}
\end{align}
\end{subequations}
with the abbreviations $\hat{\bmf}\colon \Re^{n+2} \varrightarrow \Re^{n}$, $\hat{\stg}\colon \Re^{n+3} \varrightarrow \Re$,
\begin{align}
    \hat{\bmf}(\bar{\bmZeta}, \bar{\Eta}, \bar{\vel}) & = \tfrac{1}{\bar{\vel}} \bmf(\bar{\bmZeta}, \bar{\Eta}), &
    \hat{\stg}(\bar{\bmZeta}, \bar{\Eta}, \bar{\vel}, \check{\vel}) & = \tfrac{\check{\vel}}{\bar{\vel}} \stg(\bar{\bmZeta},
    \bar{\Eta}).
    \label{eq:model:del:funcs}
\end{align}
On the domain
\begin{equation}
    \big\{(z,t) \colon z \in \spatDom,\, t > \tS + \tau(L - z,t)\big\},
    \label{eq:charSys:pde:timeSpace}
\end{equation}
the \gls{ic} $\Eta\Del(z, t) = \Eta[0](z)$ for $\varphi(z,t) = \tS$, i.e., on the boundary of
\eqref{eq:charSys:pde:timeSpace}, and
\begin{subequations}\label{eq:charSys:inout}
\begin{align}
    \ctrlIn\Del(t) & = \bmVo\bmZeta\Del(0, t) = \bmVo\bmZeta(0, t - \tau(L,t)), \label{eq:charSys:inout:ctrlIn}\\
    \bdistIn\Del(t) & = \bmV\bmZeta\Del(0, t) = \bmV\bmZeta(0, t - \tau(L,t)), \label{eq:charSys:inout:distIn}\\
    \ctrlOut\Del(t) & = \bmW\bmZeta\Del(L, t) = \bmW\bmZeta(L, t - \tau(0,t)) \label{eq:charSys:inout:out}
\end{align}
\end{subequations}
hold.
Since $\tau(0,t) = 0$ follows from \eqref{eq:charSys:delay} for $z = L$, the output is not delayed, i.e., $\ctrlOut\Del(t) = \ctrlOut(t)$, whereas
the inputs $\ctrlIn\Del(t)$ and $\bdistIn\Del(t)$ are delayed by the total transport delay $\tau(L,t)$.

\subsection{Linear case}

A linear representation of the semi-explicit \gls{pdae} system \eqref{eq:model:del} has the form
\begin{subequations}\label{eq:model:lin:del}
\begin{align}
    \pderi[z] \bmZeta\Del(z,t) & =  \tfrac{1}{\vel\Del(z,t)} \left(\bmA[1] \bmZeta\Del(z,t) + \bmB[1] \Eta\Del(z,t)\right)
    \label{eq:model:lin:del:transport}\\
    \pderi[t] \Eta\Del(z,t) & = \tfrac{\vel(t)}{\vel\Del(z,t)} \left(\bmA[2] \bmZeta\Del(z,t)+ \B[2]  \Eta\Del(z,t)\right)
    \label{eq:model:lin:del:storage}
\end{align}
\end{subequations}
with the \gls{ic} $\Eta\Del(z, t) = \Eta[0](z)$ for $\varphi(z,t) = \tS$ and the constant matrices $\bmA[1] \in \Re^{n \times n}$,
$\bmB[1] \in \Re^{n}$, and $\bmA[2]^{\trans} \in \Re^{n}$, as well as the real-valued constant $\B[2] \in \Re$.
This representation is not a direct consequence of introducing the small-signal variables and applying the Taylor approximation, since the matrices
$\bmA[1]$, $\bmB[1]$, $\bmA[2]$, and the value $\B[2]$ appear as constants.
Linearisation around a specific equilibrium would instead be expected to result in space-dependent matrices, or, in the case of linearisation around a
trajectory, in time- and space-dependent matrices.

\section{Feedforward design}
\label{sec:feedForward}

\begin{figure}[h]
    \vspace{-3ex}
    \centering
    \includegraphics[width=0.5\textwidth]{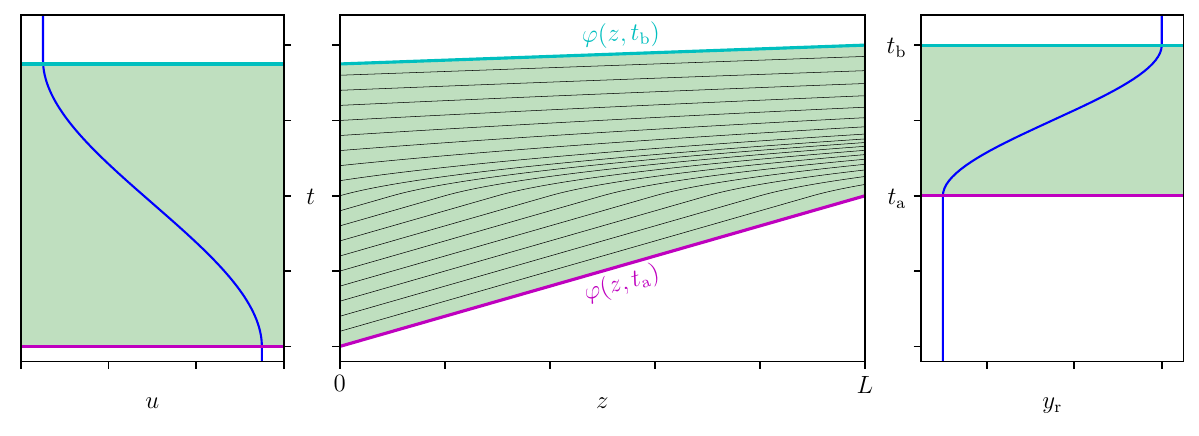}
    \caption{Consideration of the characteristic curves $\varphi(z,\tsA)$ and $\varphi(z,\tsB)$ when planning $\ctrlOut[r](t)$ on the interval
    $(\tsA,\tsB)$ for the input trajectory $\ctrlIn(t)$.}
    \label{fig:control:traj:char}
    \vspace{-4ex}
\end{figure}

For system \eqref{eq:model}, a transition between the stationary regimes $(\ctrlOut[a],\ctrlOut[b])$ is to be carried out on the interval $(\tsA,
\tsB)$.
For the realisation, a suitable reference trajectory $t\varmapsto \ctrlOut[r](t)$ is specified as
\begin{align}
    \ctrlOut[r](t) & \!=\! \begin{cases}
        \ctrlOut[a], & t \le \tsA,\\
        \ctrlOut[a] + (\ctrlOut[b] - \ctrlOut[a]) \kappa\left(\frac{t - \tsA}{\tsB - \tsA}\right), & \tsA < t < \tsB, \\
        \ctrlOut[b], & t \ge \tsB,
    \end{cases} \label{eq:control:traj}
\end{align}
using a sufficiently often continuously differentiable normalised polynomial $\kappa \colon [0,1] \varrightarrow\, [0,1]$.
Then, the input trajectory $t\varmapsto \ctrlIn(t)$ is determined by means of model inversion based on model
\eqref{eq:model:del}, using the a priori known trajectories for the disturbances $\bdistIn(t)$ and the velocity $\vel(t)$ as well as its
time derivative.
Note that the computation of the input signal $\ctrlIn\Del(t) = \ctrlIn(t - \tau(L,t))$ is shifted accordingly along the characteristic curves
$\varphi(z,\tsA)$ and $\varphi(z, \tsB)$ from the start time $\tsA$ to the end time $\tsB$.
This behaviour is illustrated in \figref{fig:control:traj:char}.
Consequently, the input $\ctrlIn(t)$ applied at time $t$ requires the reference at the future time instant $t + \hat{\tau}(L,t)$, where
$\hat{\tau}(z,t)$ denotes the delay along the forward characteristics, which is taken into account by the predictor in \secref{sec:controller}.
It should be noted that the feedforward trajectory and the state trajectories will asymptotically converge to the stationary profile
corresponding to the constant $\ctrlOut[b]$ in $\{(z,t) \mid z\in \spatDom, t >
\varphi(z,\tsB)\}$ if the internal dynamics exist and are stable.

To convey the underlying idea, the control design is first discussed on the basis of the linear variant \eqref{eq:model:lin:del}, which is comparable
to the design in \cite{Wurm2024}.
At this point, however, the state transformation of the transport process variables is omitted in order to illustrate the actual design steps more
clearly.
This is followed by a consideration of the nonlinear variant.
In each case, the discussion begins with a general consideration of the relation between the inputs and the output, before proceeding with the actual
design.

\subsection{Linear case}
For the sake of a clear presentation, the semi-explicit \gls{pdae} system \eqref{eq:model:lin:del} is simplified by assuming a constant
velocity $\vel \in \Re^{+}$.
Moreover, the diacritical mark $\square\Del$ is omitted for the sake of readability.
This yields the system
\begin{subequations}\label{eq:ff:lin}
\begin{align}
    \pderi[z] \bmZeta(z,t) & =  \hbmA[1] \bmZeta(z,t) + \hbmB[1] \Eta(z,t) \label{eq:ff:lin:transport}\\
    \pderi[t] \Eta(z,t) & = \bmA[2] \bmZeta(z,t) + \B[2] \Eta(z,t), \label{eq:ff:lin:storage}
\end{align}
\end{subequations}
which, together with its \glspl{bc} \eqref{eq:charSys:inout}, is examined with regard to its solvability using the scaled matrices $\hbmA[1] =
\nicefrac{\bmA[1]}{\vel}$ and $\hbmB[1] = \nicefrac{\bmB[1]}{\vel}$.
For this purpose, the general solution
\begin{align}
    \bmZeta(z,t) & = \uPhi(z)\bigg(\uPhi(-s) \bmZeta(s,t) + \int_{s}^{z} \uPhi(-\xi) \hbmB[1] \Eta(\xi,t) \dd \xi\bigg)
    \label{eq:ff:lin:transport:genSol}
\end{align}
of the \gls{ode} \eqref{eq:ff:lin:transport} is evaluated at $s=0$, using the transition matrix $\uPhi(z) = \exp(\hbmA[1] z)\in
\Re^{n \times n}$ and the boundary value $\bmZeta[in](t) = \bmZeta(0,t)$ for arbitrary $\Eta(\cdot,t)$.
The result
\begin{align}
    \bmZeta(z,t) & = \uPhi(z) \bmZeta[in](t) + \int_{0}^{z} \uPhi(z - \xi) \hbmB[1] \Eta(\xi,t) \dd \xi
    \label{eq:ff:lin:transport:sol}
\end{align}
is substituted into \eqref{eq:ff:lin:storage}, yielding the evolutionary \gls{ode}
\begin{multline}
    \pderi[t] \Eta(z,t) = \bmA[2] \uPhi(z) \bmZeta[in](t) \\
                        + \bmA[2] \int_{0}^{z} \uPhi(z - \xi) \hbmB[1] \Eta(\xi,t) \dd \xi + \B[2] \Eta(z,t). \label{eq:ff:lin:storage:ode}
\end{multline}
For the output $\ctrlOut(t)$, the output equation \eqref{eq:charSys:inout:out} yields
\begin{align}
    \ctrlOut(t) & = \bmW \bmZeta(L, t) \label{eq:ff:lin:outputEq} \\
         & = \bmW \uPhi(L) \bmZeta[in](t) + \bmW \int_{0}^{L} \uPhi(L - \xi) \hbmB[1] \Eta(\xi,t) \dd \xi \notag
\end{align}
when \eqref{eq:ff:lin:transport:sol} is taken into account.
Due to the decomposition $\bmZeta[in](t) = (\bmVo)^{\trans} \ctrlIn(t) + \bmV^{\trans} \bdistIn(t)$, the dependence of the term $\bmW \uPhi(L)
\bmZeta[in](t)$ on the control input can be assessed by examining only the corresponding column.
This gives the necessary and sufficient condition
\begin{align}
    \bmW \uPhi(L) (\bmVo)^{\trans} & \neq 0,
    \label{eq:ff:lin:inputCond}
\end{align}
which indicates whether the control input is directly linked to the transport process variables defining the output.
In that case, the system exhibits direct feedthrough and has a relative degree $r=0$, see, e.g., \cite{Ilchmann2016}.
This case is referred to as the \textit{coupled} case.
Otherwise, the control input acts on the output only indirectly via the storage variable, which is referred to as the \textit{uncoupled} case
and arises, e.g., if $\hbmA[1]$ is block-diagonal w.r.t. the control input and the disturbances.
To examine this case in more detail, \eqref{eq:ff:lin:outputEq} is differentiated w.r.t. time, yielding
\begin{align*}
    \dtCtrlOut(t) & = \bmW \uPhi(L) \dot{\bmZeta}_{\ttext{in}}(t) + \bmW \int_{0}^{L} \uPhi(L - \xi) \hbmB[1] \pderi[t] \Eta(\xi,t) \dd \xi.
\end{align*}
If \eqref{eq:ff:lin:inputCond} is not satisfied, the term $\bmW \uPhi(L) \dot{\bmZeta}_{\ttext{in}}(t)$ includes only the time derivatives of the
disturbances $\bdistIn(t)$.
Substituting \eqref{eq:ff:lin:storage:ode} into the latter equation and separating the inputs gives
\begin{equation}
    \hspace{-0.5ex}
    \dtCtrlOut(t) = D_1 \bdistIn(t) + D_2 \dot{\bdistIn}(t)
                  + (\opP[1] + \opP[2]) \Eta(\cdot, t)
                  + Q \ctrlIn(t)
    \label{eq:ff:lin:dty:uncoupled}
\end{equation}
using the parameters
\begin{subequations}\label{eq:ff:lin:dty:uncoupled:params}
\begin{align}
    D_1 & = \bmW \int_{0}^{L} \uPhi(L - \xi) \hbmB[1] \bmA[2] \uPhi(\xi) \bmV^{\trans} \dd \xi, \\
    D_2 & = \bmW \uPhi(L) \bmV^{\trans}, \\
    Q & = \bmW \int_{0}^{L} \uPhi(L - \xi) \hbmB[1] \bmA[2] \uPhi(\xi) (\bmVo)^{\trans} \dd \xi,
\end{align}
and the Fredholm integral operators
\begin{align}
    \opP[1] \bx & = \bmW \int_{0}^{L} \!\uPhi(L - \xi) \hbmB[1] \bmA[2] \int_{0}^{\xi} \!\uPhi(\xi - \vartheta) \hbmB[1] \bx(\vartheta) \dd \vartheta \dd \xi, \\
    \opP[2] \bx & = \bmW \int_{0}^{L} \!\uPhi(L - \xi) \hbmB[1] \B[2] \bx(\xi) \dd \xi.
\end{align}
\end{subequations}
It becomes clear that, if \eqref{eq:ff:lin:inputCond} is violated, the system has relative degree $r=1$ if and only if
\begin{align}
    Q & \neq 0
    \label{eq:ff:lin:inputCondRel}
\end{align}
holds.
Since $Q$ is the integral of the product of the scalar functions $a(z) = \bmW \uPhi(L-z) \hbmB[1]$ and $b(z) = \bmA[2] \uPhi(z)
(\bmVo)^{\trans}$, a necessary condition for \eqref{eq:ff:lin:inputCondRel} is that neither $a(z)$ nor $b(z)$ vanishes identically on $\spatDom$,
i.e., the output depends on the storage variable and the control input acts on it.
However, these structural conditions are not sufficient, since the integral may vanish for a specific choice of $L$, which results in a
relative degree greater than $1$.
Furthermore, if $b(z) \equiv 0$, i.e., the control input does not act on the storage variable at all, the output cannot be influenced by the
control input.

The following lemma summarises the case considered in this contribution.
\begin{lemma}
    When
    \begin{align*}
        \bmW \uPhi(L) (\bmVo)^{\trans} & \neq 0
    \end{align*}
    holds, the system has relative degree $r=0$ and is referred to as the \textit{coupled} case.
    Otherwise, when
    \begin{align*}
        Q = \bmW \int_{0}^{L} \uPhi(L - \xi) \hbmB[$\mathrm{1}$] \bmA[$\mathrm{2}$] \uPhi(\xi) (\bmVo)^{\trans} \mathrm{d} \xi & \neq 0
    \end{align*}
    holds, the system has relative degree $r=1$ and is referred to as the \textit{uncoupled} case.
\end{lemma}
This is similar to the property of input-output controllability for the linear, finite-dimensional case, as described, e.g., in \cite{Lunze2016}.
In the subsequent sections, the \textit{coupled} and the \textit{uncoupled} cases are analysed with regard to the design of the feedforward law.

\subsubsection{Coupled case}
\label{sec:ff:lin:coupled}
This case satisfies the condition \eqref{eq:ff:lin:inputCond}, which implies relative degree $r=0$.
There are no differentiability conditions on $\ctrlOut[r](t)$, so the polynomial $\kappa(\tau)$ in \eqref{eq:control:traj} can be chosen as
$\kappa(\tau) = \tau$.
The control signal $\ctrlIn(t)$ is the result of \eqref{eq:charSys:inout:ctrlIn}, wherein $\bmZeta[r](0,t)$ follows from the
linear \gls{bvp}:
\begin{subequations}\label{eq:ff:lin:coupled:bvp}
\begin{align}
    \bar{\bmZeta}_{\ttext{r}}' & =  \hbmA[1] \bar{\bmZeta}_{\ttext{r}} + \hbmB[1] \bar{\Eta}_{\ttext{r}}, \\
    \bm{S}_{0} \bar{\bmZeta}_{\ttext{r}}(0) & + \bm{S}_{L} \bar{\bmZeta}_{\ttext{r}}(L) = \bar{\bmInputs} \label{eq:ff:lin:coupled:bvp:bcs}
\end{align}
including
\begin{align}
    \bm{S}_{0} & = \begin{pmatrix} \bmV \\ 0 \end{pmatrix}, \
    \bm{S}_{L} = \begin{pmatrix}  \bm{0}_{(n-1)\times n} \ \\ \bmW \end{pmatrix}, \
    \bar{\bmInputs} = \begin{pmatrix} \bar{\bdistIn} \\ \bar{\ctrlOut}_{\ttext{r}} \end{pmatrix},
    \label{eq:ff:lin:coupled:bvp:params}
\end{align}
\end{subequations}
which combines \eqref{eq:ff:lin:transport} with the \glspl{bc} in the form of the reference output $\bar{\ctrlOut}_{\ttext{r}}$ and the a priori
known disturbances $\bar{\bdistIn}$.

Evaluating \eqref{eq:ff:lin:transport:genSol} first for $s = 0$ and multiplying the result by $\bm{S}_0 \uPhi(-z)$, then for $s = L$
and multiplying by $\bm{S}_{L}\uPhi(L)\uPhi(-z)$, allows \eqref{eq:ff:lin:coupled:bvp:bcs} to be substituted into the sum of the two resulting
equations, yielding the solution of \eqref{eq:ff:lin:coupled:bvp}:
\begin{align}
    \bar{\bmZeta}_{\ttext{r}}(z) & = \hat{\bmG}(z) \bar{\bmInputs} + \int_{0}^{L} \bmG(z,\xi) \hbmB[1] \bar{\Eta}_{\ttext{r}}(\xi) \dd \xi.
    \label{eq:ff:lin:coupled:bvp:sol}
\end{align}
In the latter, the Green's matrix $\bmG(z,\xi) \in \Re^{n \times n}$, see, e.g., \cite{Whyburn1942},
\begin{align*}
    \bmG(z,\xi) & = \begin{cases}
        \hat{\bmG}(z) \bm{S}_0 \uPhi(-\xi) & 0 \le \xi \le z,\\
        -\hat{\bmG}(z) \bm{S}_L \uPhi(L) \uPhi(-\xi) & z < \xi \le L
    \end{cases}
\end{align*}
and the matrix $\hat{\bmG}(z) = \uPhi(z) \bmO^{-1} \in \Re^{n \times n}$ appear, including the boundary matrix $\bmO = \bm{S}_0 + \bm{S}_L
\uPhi(L) \in \Re^{n \times n}$, which is invertible by virtue of condition \eqref{eq:ff:lin:inputCond}.

Finally, the control input can be expressed as
\begin{align*}
    \ctrlIn(t) & = \bmVo \bmZeta[r](0,t) \\
                   & = \bmVo \hat{\bmG}(0) \bmInputs(t) + \bmVo \int_{0}^{L} \bmG(0,\xi) \hbmB[1] \Eta[r](\xi,t) \dd \xi, \notag
\end{align*}
using \eqref{eq:ff:lin:coupled:bvp:sol}, wherein $\Eta[r](\cdot,t)$ appears, which is a result of the integrated internal dynamics.
This follows from substituting \eqref{eq:ff:lin:coupled:bvp:sol} into \eqref{eq:ff:lin:storage} and yields the abstract
evolutionary \gls{ode}
\begin{align}
    \dot{x}(t) & = \opA x(t) + \opB \bmInputs(t), \
    t > \tS, \
    x(\tS) = x_0 \in X
    \label{eq:ff:lin:coupled:abstractOde}
\end{align}
by introducing the state $x(\cdot,t) = \Eta[r](\cdot, t) \in X$ and the bounded linear system operator $\opA \colon X \varrightarrow X$, $\opA =
\opA[c,c] + \B[2]$, with $(\opA[c,c] \bx)(z) = \bmA[2] \int_{0}^{L} \bmG(z,\xi) \hbmB[1] \bx(\xi) \dd \xi$ as well as the bounded linear input
operator $\opB \colon \Re^{n} \varrightarrow X$, $\opB = \bmA[2] \hat{\bmG}$, on the state space $X = \Cont[$0$](\spatDom)$.
Since $\opA$ is bounded, it generates a uniformly continuous semigroup, so that \eqref{eq:ff:lin:coupled:abstractOde} is well-posed.
Stability of the internal dynamics is a necessary requirement and is investigated in the following by analysing the spectrum of the operator $\opA$.
As an integral operator with piecewise continuous kernel, $\opA[c,c]$ is compact, and $\opA = \B[2] \opI + \opA[c,c]$ is a compact perturbation of
a multiple of the identity.
Consequently, see, e.g., \cite{Heuser1992}, the spectrum of $\opA$ consists of at most countably many eigenvalues $s_i \neq \B[2]$ of finite
multiplicity, which can only accumulate at $\B[2]$, together with the point $\B[2]$ itself, which belongs to the spectrum since $X$ is
infinite-dimensional.
The eigenvalues follow from the eigenvalue problem $(s \opI - \opA) \bx = 0$, and the spectrum is given by
\begin{equation}
    \spec(\opA) = \{s\in\Co \mid \exists \bx\neq 0\colon (s \opI -\opA)\bx =0\} \cup \{\B[2]\}.
    \label{eq:ff:lin:coupled:spec}
\end{equation}
Since the operator $\opA$ is bounded, the \textit{Spectrum Determined Growth Assumption} holds, i.e., the internal dynamics are exponentially
stable if and only if $\B[2] < 0$ and $\text{Re}(s_i) < 0$ for all eigenvalues.
If infinitely many eigenvalues exist, they lie arbitrarily close to $\B[2]$ for sufficiently large $i \in \N$.
Nevertheless, even for $\B[2] < 0$, a finite number of eigenvalues may still lie in the right half-plane.
Consequently, a general statement regarding stability cannot be made without a detailed analysis, and only the conservative estimate $\text{Re}(s_i)
\le \B[2] + \varrho(\opA[c,c])$ can be given, where $\varrho(\opA[c,c])$ denotes the spectral radius of $\opA[c,c]$ on $X$.

\subsubsection{Uncoupled case}
\label{sec:ff:lin:uncoupled}
When considering this case, condition \eqref{eq:ff:lin:inputCond} is not satisfied, but condition \eqref{eq:ff:lin:inputCondRel} is.
Consequently, the system has relative degree $r=1$.
Hence, $\ctrlOut[r](t)$ must now be at least once continuously differentiable, and a third-degree polynomial may be chosen for $\kappa(\tau)$
in \eqref{eq:control:traj}.
The polynomial $\kappa(\tau) = 3 \tau^2 - 2 \tau^3$ satisfies precisely these requirements.
The control signal $\ctrlIn(t)$ follows directly from \eqref{eq:ff:lin:dty:uncoupled} as
\begin{align}
    \ctrlIn(t) & = \tfrac{\dtCtrlOut[r](t) - D_1 \bdistIn(t) - D_2 \dot{\bdistIn}(t)
                     - (\opP[1] + \opP[2]) \Eta[r](\cdot, t)}{Q} \label{eq:ff:lin:uncoupled:input}
\end{align}
in which, in addition to the disturbances $\bdistIn(t)$ and their time derivatives, $\Eta[r](\cdot, t)$ also appears.
The latter is a result of integrating the internal dynamics over time.
To obtain it, \eqref{eq:ff:lin:uncoupled:input} is substituted into \eqref{eq:ff:lin:storage:ode}, which gives the evolutionary \gls{ode}
\begin{multline*}
    \pderi[t] \Eta[r](z,t) = S(z) \dtCtrlOut[r](t) + \hat{D}_1(z) \bdistIn(t) + \hat{D}_2(z) \dot{\bdistIn}(t) \\
                           + ((\opP[3] + \B[2])\Eta[r])(z, t).
\end{multline*}
Therein, the spatially dependent parameters
\begin{align*}
    \hat{D}_1(z) & = \bmA[2] \uPhi(z) \bmV^{\trans} - \tfrac{\bmA[2] \uPhi(z) (\bmVo)^{\trans} D_1}{Q}, \\
    \hat{D}_2(z) & = -\tfrac{\bmA[2] \uPhi(z) (\bmVo)^{\trans} D_2}{Q}, \quad
    S(z) = \tfrac{\bmA[2] \uPhi(z) (\bmVo)^{\trans}}{Q},
\end{align*}
and the operator
\begin{align*}
    (\opP[3] \bx)(z) & = (\opA[u,c] \bx)(z) - S(z) (\opP[1] + \opP[2])\bx
\end{align*}
are introduced using the parameters and operators from \eqref{eq:ff:lin:dty:uncoupled:params}, as well as the Volterra integral operator $(\opA[u,c]
\bx)(z) = \bmA[2] \int_{0}^{z} \uPhi(z - \xi) \hbmB[1] \bx(\xi) \dd \xi$.
The latter equation still contains derivatives of the reference $\ctrlOut[r](t)$ and the disturbances $\bdistIn(t)$.
To eliminate these, the state transformation
\begin{align}
    x(\cdot,t) & = \Eta[r](\cdot, t) - S(\cdot)\ctrlOut[r](t) - \hat{D}_2(\cdot) \bdistIn(t) \in X
    \label{eq:ff:lin:uncoupled:state}
\end{align}
is introduced, which leads to an abstract state-space representation of the internal dynamics of the form \eqref{eq:ff:lin:coupled:abstractOde}.
The bounded linear system operator $\opA \colon X \varrightarrow X$ now appearing therein is given by $\opA = \opP[3] + \B[2]$.
The likewise bounded linear input operator $\opB \colon \Re^{n} \varrightarrow X$ for the inputs $\bar{\bmInputs} = (\bar{\bdistIn},
\bar{\ctrlOut}_{\ttext{r}})^{\trans}$ follows analogously from $\hat{D}_1$, $\hat{D}_2$, and $S$.
To specify the state space $X$, the output equation \eqref{eq:ff:lin:outputEq} with $\ctrlOut(t) = \ctrlOut[r](t)$ must be satisfied in the new
coordinates.
Substituting \eqref{eq:ff:lin:uncoupled:state} yields this condition as $\bmW \int_{0}^{L} \uPhi(L - \xi) \hbmB[1] x(\xi,t) \dd \xi = 0$
and the state space is therefore obtained as $X = \{ x \in \Cont[$0$](\spatDom) \colon \opC x = 0 \}$ with $\opC x = \bmW \int_{0}^{L} \uPhi(L - \xi)
\hbmB[1] x(\xi) \dd \xi$.
This space is invariant under $\opA$ and $\opB$:
By the definitions of $\opP[1]$, $\opP[2]$, and $Q$, one obtains $\opC \opA[u,c] \bx = \opP[1] \bx$, $\opC (\B[2] \bx) = \opP[2] \bx$, and
$\opC S = 1$, so that $\opC \opA \bx = \opP[1] \bx - (\opP[1] + \opP[2]) \bx + \opP[2] \bx = 0$ holds for all $\bx$; analogously,
$\opC \opB \bar{\bmInputs} = 0$ follows.
This shows that, in contrast to the \textit{coupled} case, in which the internal dynamics corresponded to the state space of $\Eta(\cdot, t)$, the
state space of the internal dynamics must now, in this case, be restricted by one condition, induced by the relative degree $r=1$.
This is comparable to the finite-dimensional case, in which the internal dynamics have dimension $n - r$, where $n$ denotes the state dimension; see,
for instance, \cite{Khalil2002}.

As in the \textit{coupled} case, it is essential that the internal dynamics are stable, which can likewise be examined by analysing the spectrum
of the operator $\opA$.
The operator $\opP[3]$ consists of the Volterra integral operator $\opA[u,c]$ and a term composed of the Fredholm integral operators $\opP[1]$
and $\opP[2]$, all of which are compact due to their piecewise continuous kernels.
Hence, $\opA = \B[2] \opI + \opP[3]$ is again a compact perturbation of a multiple of the identity on the infinite-dimensional space $X$, and its
spectrum is of the form \eqref{eq:ff:lin:coupled:spec}, see \cite{Heuser1992}.
Due to the boundedness of $\opA$, the \textit{Spectrum Determined Growth Assumption} holds, and the same stability conditions as in the
\textit{coupled} case apply, with the conservative estimate $\text{Re}(s_i) \le \B[2] + \varrho(\opP[3])$.

\subsection{Nonlinear case}
As in the linear case, the investigation of the feedforward design is based on a simplified version of the semi-explicit \gls{pdae} system
\eqref{eq:model:del}, which can be written as
\begin{subequations}\label{eq:ff:nonlin}
\begin{align}
    \pderi[z] \bmZeta(z,t) & =  \tfrac{1}{\vel} \bmf(\bmZeta(z,t), \Eta(z,t)) \label{eq:ff:nonlin:transport}\\
    \pderi[t] \Eta(z,t) & = \stg(\bmZeta(z,t), \Eta(z,t)) \label{eq:ff:nonlin:storage}
\end{align}
\end{subequations}
with constant velocity $\vel$.
The functions $\bmf$ and $\stg$ in \eqref{eq:ff:nonlin} are assumed to be continuously differentiable, i.e.,
$\bmf \in \Cont[$1$](\Re^{n+1}, \Re^{n})$ and $\stg \in \Cont[$1$](\Re^{n+1}, \Re)$.
Moreover, the \gls{ode} \eqref{eq:ff:nonlin:transport} is assumed to be forward-complete w.r.t. $z$, i.e., a unique solution $\bmZeta(z,\cdot)$
exists on the spatial domain $\spatDom$ for every \gls{bc} $\bmZeta(0, t) = \bmZeta[in](t) \in \Re^{n}$ and every
profile $\Eta(\cdot, t) \in \Cont[$0$](\spatDom)$, see \cite{Sontag1998, Khalil2002}.
Note that forward completeness excludes finite escape time.

Under these conditions, the general solution of the nonlinear \gls{ode} \eqref{eq:ff:nonlin:transport} can be parameterised as
\begin{align}
    \bar{\bmZeta}(z) & = \odesFlow[$\hat{\bmf}$](z, \bar{\ctrlIn}, \bar{\bdistIn}, \chi_{[0,z]}\bar{\Eta}),
    \label{eq:ff:nonlin:transport:sol}
\end{align}
using the flow $\odesFlow[$\hat{\bmf}$] \colon \spatDom \times \Re^{n} \times \Cont[$0$](\spatDom) \varrightarrow \Re^{n}$ w.r.t. the control input
$\bar{\ctrlIn}$ and the disturbance $\bar{\bdistIn}$.
Here, the characteristic function $\chi_{[0,z]}$ indicates that $\odesFlow[$\hat{\bmf}$]$ depends on $\bar{\Eta} \in \Cont[$0$](\spatDom)$ only via
its values on the interval $[0,z]$.
Substituting \eqref{eq:ff:nonlin:transport:sol} into \eqref{eq:ff:nonlin:storage} yields the nonlinear evolutionary \gls{ode}
\begin{align}
    \pderi[t] \Eta(z,t) & = \breve{\stg}(z, \ctrlIn(t), \bdistIn(t), \Eta(\cdot, t))
    \label{eq:ff:nonlin:storage:ode}
\end{align}
with the function $\breve{\stg} \colon \spatDom \times \Re^{n} \times \Cont[$0$](\spatDom) \varrightarrow \Re$,
\begin{align*}
    \breve{\stg}(z, \bar{\ctrlIn}, \bar{\bdistIn}, \bar{\Eta}) = \stg(\odesFlow[$\hat{\bmf}$](z, \bar{\ctrlIn}, \bar{\bdistIn}, \chi_{[0,z]}\bar{\Eta}), \bar{\Eta}(z)).
\end{align*}
Furthermore, the output equation \eqref{eq:charSys:inout:out} can be expressed as
\begin{align}
    \ctrlOut(t) & = \bmW \bmZeta(L, t)
                = \bmW \odesFlow[$\hat{\bmf}, L$](t)
                \label{eq:ff:nonlin:output}
\end{align}
by virtue of \eqref{eq:ff:nonlin:transport:sol}, using the abbreviation $\odesFlow[$\hat{\bmf}, L$](t) = \odesFlow[$\hat{\bmf}$](L, \ctrlIn(t),
\bdistIn(t), \Eta(\cdot,t))$.
To determine whether the output can be directly influenced by the input $\ctrlIn(t)$, the condition
\begin{align}
    \bmW \pderi[\ctrlIn] \odesFlow[$\hat{\bmf}, L$](t) & \neq 0,
    \label{eq:ff:nonlin:inputCond}
\end{align}
is examined along specific trajectories of $\ctrlIn(t)$, $\bdistIn(t)$, and $\Eta(\cdot,t)$, in analogy to condition
\eqref{eq:ff:lin:inputCond} in the linear case.
If \eqref{eq:ff:nonlin:inputCond} holds, it remains valid in a neighbourhood of the given trajectories, and the output $\ctrlOut(t)$ is thus directly
reachable from the input $\ctrlIn(t)$.
Hence, the system has relative degree $r=0$ and corresponds to the \textit{coupled} case.
If \eqref{eq:ff:nonlin:inputCond} is not satisfied, i.e., $\bmW \pderi[\ctrlIn] \odesFlow[$\hat{\bmf}, L$](t) \equiv 0$, the term involving
$\dot{\ctrlIn}(t)$ vanishes in the time derivative of the output, and substituting \eqref{eq:ff:nonlin:storage:ode} yields
\begin{multline}
    \dtCtrlOut(t) = \bmW \pderi[\bdistIn]\odesFlow[$\hat{\bmf}, L$](t)\, \dtbdistIn(t) \\
    + \bmW \pderi[\Eta]\odesFlow[$\hat{\bmf}, L$](t)\, \breve{\stg}(\cdot, \ctrlIn(t), \bdistIn(t), \Eta(\cdot, t)), \label{eq:ff:nonlin:dty:uncoupled}
\end{multline}
where, for notational simplicity, the partial derivatives are to be understood as Gâteaux derivatives.
If the condition
\begin{align}
    \bmW \pderi[\Eta]\odesFlow[$\hat{\bmf}, L$](t)\, \pderi[\ctrlIn] \breve{\stg}(\cdot, \ctrlIn(t), \bdistIn(t), \Eta(\cdot, t))
    & \neq 0, \label{eq:ff:nonlin:inputCondRel}
\end{align}
is satisfied for a specific $\ctrlIn(t)$, $\bdistIn(t)$, and a trajectory of $\Eta(\cdot, t)$, then \eqref{eq:ff:nonlin:dty:uncoupled} can be
solved for $\ctrlIn(t)$ by virtue of the implicit function theorem \cite{Zeidler1995}, although this holds only in a neighbourhood of the given
values.
In this case, the system has relative degree $r = 1$ and corresponds to the \textit{uncoupled} case, analogous to the linear setting.
If neither \eqref{eq:ff:nonlin:inputCondRel} nor \eqref{eq:ff:nonlin:inputCond} is satisfied, the relative degree is greater than one or the
output cannot be influenced by the control input at all.
In what follows, analogous to the linear representation, the \textit{coupled} and \textit{uncoupled} cases are treated separately for the feedforward design.

\subsubsection{Coupled case}
\label{sec:ff:nonlin:coupled}
The condition \eqref{eq:ff:nonlin:inputCond} is satisfied, which implies a relative degree $r=0$ and, similar to the linear variant, imposes no
differentiability conditions on $\ctrlOut[r](t)$.
Therefore, the same approach as in \secref{sec:ff:lin:coupled} can be used, both for $\kappa(\tau)$ and for the derivation described therein for the
control input $\ctrlIn(t)$ based on \eqref{eq:charSys:inout:ctrlIn}.
The boundary value $\bmZeta[r](0,t)$ follows from the nonlinear \gls{bvp}:
\begin{align}
    \bar{\bmZeta}_{\ttext{r}}' & = \tfrac{1}{\vel} \bmf(\bar{\bmZeta}_{\ttext{r}}, \bar{\Eta}_{\ttext{r}}), &
    \bm{S}_{0} \bar{\bmZeta}_{\ttext{r}}(0) + \bm{S}_{L} \bar{\bmZeta}_{\ttext{r}}(L) & = \bar{\bmInputs}
    \label{eq:ff:nonlin:coupled:bvp}
\end{align}
with the parameters given in \eqref{eq:ff:lin:coupled:bvp:params} and the input $\bar{\bmInputs} = (\bar{\bdistIn},
\bar{\ctrlOut}_{\ttext{r}})^{\trans}\in \Re^{n}$.
The solution of the latter equation is given by
\begin{align}
    \bar{\bmZeta}_{\ttext{r}}(z) & = \hOdesFlow[$\hat{\bmf}$](z, \bar{\bmInputs}, \chi_{[0,L]}\bar{\Eta}_{\ttext{r}}),
    \label{eq:ff:nonlin:coupled:bvp:sol}
\end{align}
using the map $\hOdesFlow[$\hat{\bmf}$] \colon \spatDom \times \Re^{n} \times \Cont[$0$](\spatDom) \varrightarrow \Re^{n}$ and the characteristic function
$\chi$.
The existence of this solution is not guaranteed in general, but can be ensured locally.
Using the parameterisation \eqref{eq:ff:nonlin:transport:sol}, the \glspl{bc} associated with the disturbances are satisfied by
construction, and \eqref{eq:ff:nonlin:coupled:bvp} reduces to the scalar equation $\bmW \odesFlow[$\hat{\bmf}$](L, \bar{\ctrlIn}, \bar{\bdistIn},
\bar{\Eta}_{\ttext{r}}) = \bar{\ctrlOut}_{\ttext{r}}$ for the unknown $\bar{\ctrlIn} = \bmVo \bar{\bmZeta}_{\ttext{r}}(0)$, whose derivative
w.r.t. $\bar{\ctrlIn}$ does not vanish due to condition \eqref{eq:ff:nonlin:inputCond}.
Hence, by the implicit function theorem \cite{Zeidler1995}, a solution for given values, e.g., a stationary regime, can be uniquely
continued to a neighbourhood of these values, in which $\hOdesFlow[$\hat{\bmf}$]$ is well-defined and continuously differentiable.
In the linear case, this corresponds to the invertibility of the boundary matrix $\bm{O}$ by virtue of condition \eqref{eq:ff:lin:inputCond}.
Finally, the control input $\ctrlIn(t)$ can be written as
\begin{align*}
    \ctrlIn(t) & = \bmVo \bmZeta[r](0,t) = \bmVo \hOdesFlow[$\hat{\bmf}$](0, \bmInputs(t), \chi_{[0,L]}\Eta[r](\cdot,t)),
\end{align*}
wherein the state of the internal dynamics $\Eta[r](\cdot,t)$ appears.
The internal dynamics follow from \eqref{eq:ff:nonlin:storage} together with \eqref{eq:ff:nonlin:coupled:bvp:sol} and can be expressed as the nonlinear
abstract evolutionary \gls{ode}
\begin{align}
    \dot{x}(t) & = \check{\stg}(x(t), \bmInputs(t)), &
    t & > \tS, &
    x(\tS) & = x_0 \in X
    \label{eq:ff:nonlin:coupled:abstractOde}
\end{align}
with the map $\check{\stg} \colon X \times \Re^{n} \varrightarrow X$, defined pointwise by
\begin{align*}
    \big(\check{\stg}(\bar{x}, \bar{\bmInputs})\big)(z) & = \stg\big(\hOdesFlow[$\hat{\bmf}$](z, \bar{\bmInputs}, \chi_{[0,L]}\bar{x}), \bar{x}(z)\big),
    \quad z \in \spatDom,
\end{align*}
using the state $x(\cdot, t) = \Eta[r](\cdot, t) \in X$ on the state space $X = \Cont[$0$](\spatDom)$.
In contrast to the space of square-integrable functions, superposition operators such as $\check{\stg}(\cdot, \bar{\bmInputs})$ are
continuously differentiable on this space, which is required for the following stability analysis.
Here, too, it is important that the internal dynamics are stable.
To this end, the principle of linearised stability can be applied, see, e.g., \cite{Daleckii1974}.
For constant inputs $\bar{\bmInputs}$, let $x^{\star} \in X$ be an equilibrium of \eqref{eq:ff:nonlin:coupled:abstractOde}, i.e.,
$\check{\stg}(x^{\star}, \bar{\bmInputs}) = 0$.
Since $\bmf$ and $\stg$ are continuously differentiable, $\check{\stg}(\cdot, \bar{\bmInputs})$ is continuously differentiable on $X$, and its
derivative $\pderi[\bar{x}] \check{\stg}(\bar{x}, \bar{\bmInputs})\big|_{\bar{x}=x^{\star}}$ is a bounded linear operator.
The equilibrium is locally exponentially stable if the spectrum of this operator lies in the open left half-plane, and unstable if it contains a
point with positive real part.
The linearisation of \eqref{eq:ff:nonlin:coupled:abstractOde} around $x^{\star}$ yields an abstract \gls{ode} of the form
\eqref{eq:ff:lin:coupled:abstractOde}, but with spatially varying coefficients, since the stationary profiles depend on $z$.
The analysis of the spectrum in \secref{sec:ff:lin:coupled} would therefore have to be extended to this case, which remains an open problem.

\subsubsection{Uncoupled case}
\label{sec:ff:nonlin:uncoupled}
Unlike the \textit{coupled} case, condition \eqref{eq:ff:nonlin:inputCond} is not satisfied; instead, condition \eqref{eq:ff:nonlin:inputCondRel} holds.
The system has relative degree $r=1$.
Similar to the linear representation in \secref{sec:ff:lin:uncoupled}, $\ctrlOut[r](t)$ must now be at least once continuously differentiable, and the
same polynomial can be chosen for $\kappa(\tau)$.
To determine the control input $\ctrlIn(t)$, \eqref{eq:ff:nonlin:dty:uncoupled} is taken as the starting point.
Since the derivative w.r.t. $\bar{\ctrlIn}$ in \eqref{eq:ff:nonlin:inputCondRel} does not vanish, a function $h_{\ttext{u}} \colon \Re^{2\, n-1}
\times \Cont[$0$](\spatDom) \varrightarrow \Re$ can be found for specific values of $\dtCtrlOut[r](t)$, $\bdistIn(t)$, $\dtbdistIn(t)$, and a
trajectory of $\Eta[r](\cdot, t)$, such that, in a small neighbourhood around these, a unique solution for the control input can be computed as
\begin{align}
    \ctrlIn(t) & = h_{\ttext{u}}(\dtCtrlOut[r](t), \bdistIn(t), \dtbdistIn(t), \Eta[r](\cdot,t)),
    \label{eq:ff:nonlin:uncoupled:input}
\end{align}
where the existence and uniqueness of $h_{\ttext{u}}$ follow from the implicit function theorem.
To determine the state $\Eta[r](\cdot, t)$, the same approach as in the linear \textit{uncoupled} case is followed: substituting
\eqref{eq:ff:nonlin:uncoupled:input} into \eqref{eq:ff:nonlin:storage:ode} yields a nonlinear evolutionary \gls{ode} for $\Eta[r](\cdot, t)$, which
represents the complete dynamics of the system.
To deduce the internal dynamics, a state transformation, analogous to the linear case, would have to be introduced by means of which the output and
disturbance derivatives can be eliminated; however, this requires knowledge of the structure of $\breve{\stg}$, for which no general statements can be
made at this point.

If the internal dynamics are known, they must likewise be stable.
This can be analysed, as in the \textit{coupled} case, by means of the principle of linearised stability.
One obtains a linear abstract \gls{ode} of the form \eqref{eq:ff:lin:coupled:abstractOde} with a system operator as in
\secref{sec:ff:lin:uncoupled}, whose spectrum must then be examined.

\section{Controller design}
\label{sec:controller}

\begin{figure}[t]
    \vspace{-2ex}
	\begin{adjustbox}{width=0.5\textwidth}
	    \begin{tikzpicture}[auto, >=latex']
    \node [block, minimum width=2.6cm] (traj) {
        \eqref{eq:control:traj}
    };
    \node[boxtitle, right=10pt] at (traj.north west) {Traj.-gen.};

    \node [sum, node distance=1.875cm, below=of traj] (ctrlSum) {};

    \node [block, minimum width=2.6cm, node distance=1cm, right=of ctrlSum] (ctrl) {
        \eqref{eq:control:w}
    };
    \node[boxtitle, right=10pt] at (ctrl.north west) {Controller};

    \node [block, minimum width=2.6cm, node distance=2cm, right=of ctrl] (ff) {
        \eqref{eq:control:inverse}
    };
    \node[boxtitle, right=10pt] at (ff.north west) {Inv. Model};

    \node [branch, node distance=1cm, right=of ff] (bIn) {};

    \node [block, minimum width=2.6cm, node distance=1cm, right=of bIn] (sys) {
        \eqref{eq:model}
    };
    \node[boxtitle, right=10pt] at (sys.north west) {System};

    \node [block, minimum width=2.6cm, node distance=1cm, below=of ff] (pred) {
        \eqref{eq:control:predictor}
    };
    \node[boxtitle, right=10pt] at (pred.north west) {Predictor};

    \node[branch, node distance=2cm, above=of sys] (bdist) {};
    \node[coord, node distance=1cm, above=of bdist] (dist) {};

    \draw[->] (traj) -- node [right] {$\ctrlOut[r]\Pred(t)$} node[left, yshift=-4.75ex] {$-$} (ctrlSum);
    \draw[->] (ctrlSum) -- (ctrl);
    \draw[->] (ctrl.345) -- node [above] {$\bar{w}(t)$} (ff.195);
    \draw[-] (ff) -- (bIn) node [above] {$\ctrlIn(t)\vphantom{\ctrlIn[C](t)}$};
    \draw[->] (bIn) -- (sys);
    \draw[->] (bIn) |- (pred.15);
    \draw[->] (pred) -| node [above, xshift=3.75ex] {$\breve{\ctrlOut}\Pred(t)$} (ctrlSum);
    \draw[->] (sys) |- node [above, xshift=-3.75ex] {$\ctrlOut[m](t)$}(pred.345);
    \draw[->] (traj) -- node [above] {$\dtCtrlOut[r]\Pred(t)$} +(4.75, 0) |- (ff.165);

    \draw[->, violet] (dist) -- node [left, yshift=2ex] {$\bdistIn(t)\vphantom{\dot{\bdistIn}}$}
                              node [right, yshift=2ex] {$\dot{\bdistIn}(t)\vphantom{\dot{\bdistIn}}$}
                              node [left, yshift=-1ex] {$\vel(t)\vphantom{\dot{\bdistIn}}$}
                              node [right, yshift=-1ex] {$\dot{\vel}(t)\vphantom{\dot{\bdistIn}}$}
                        (bdist) -- ([yshift=1.4ex]sys.90);
    \draw[->, violet] (bdist) -| ([yshift=1.4ex]ff.90);
    \draw[->, violet] (ff.215) -- ([yshift=1.4ex]pred.145);
    \draw[<-] (ff.325) -- node [right] {$\hat{\Eta}\Pred(\cdot,t)$} (pred.35);
\end{tikzpicture}
 	\end{adjustbox}
    \caption{General control strategy for the \textit{uncoupled} case.}
    \label{fig:ctrl:overview}
    \vspace{-3ex}
\end{figure}
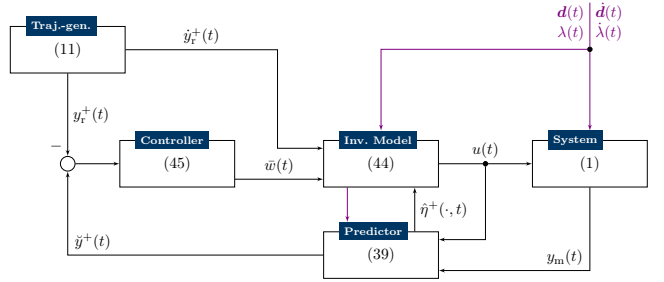

To compensate for unknown disturbances, parameter uncertainties, and errors arising from the spatial discretisation of the equations required for
implementation, a controller is incorporated into the scheme by extending the model inversion, as shown in \figref{fig:ctrl:overview}.
In practice, only a point-wise measurement of the output variable $\ctrlOut[m](\cdot) = \bmW \bmZeta(L, \cdot)$ is available.
Since transport delays occur in the system due to convection processes, these must be taken into account in the design.
\begin{figure}[b]
    \vspace{-3ex}
    \centering
    \includegraphics[width=0.5\textwidth]{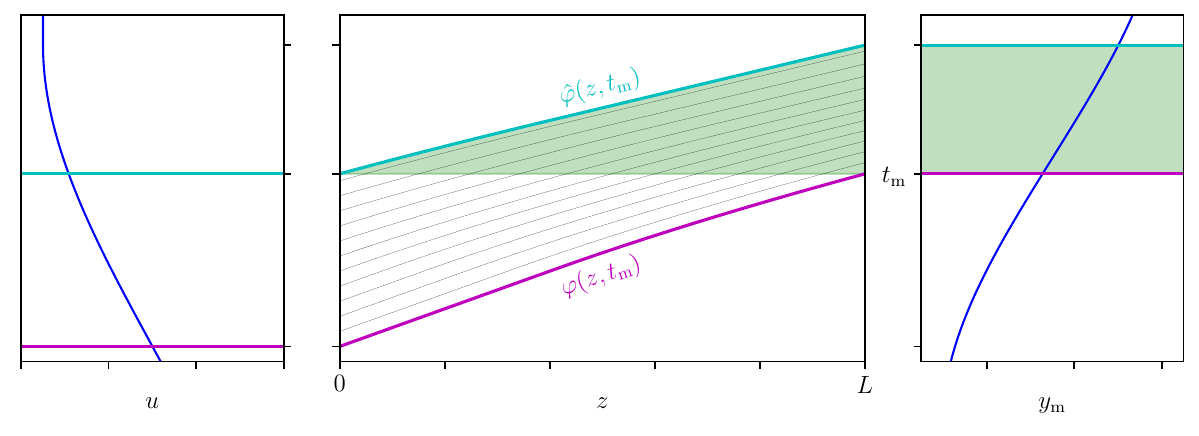}
    \caption{Representation of the temporal influence of the input variable $\ctrlIn(t)$ on the measurement $\ctrlOut[m](\tM)$ based on the
    characteristic curves $\varphi(z, \tM)$ and $\hat{\varphi}(z, \tM)$.}
    \label{fig:control:ctrl:char}
    \vspace{-4ex}
\end{figure}

\subsection{Predictor}

The control input $\ctrlIn(t)$ applied at time $t$ only becomes visible at the output $\ctrlOut\Pred(t) = \ctrlOut(t + \hat{\tau}(L,t))$ after the
transport delay $\hat{\tau}(L,t)$; see \figref{fig:control:ctrl:char}.
In contrast to the backward characteristics $\varphi(z,t)$ used in \secref{sec:problem}, the forward characteristics $\hat{\varphi}(z,t) = t +
\hat{\tau}(z,t)$ are relevant here, where the spatially and temporally dependent transport delay $\hat{\tau}(z,t)$ follows from
\begin{align}
    \pderi[z] \hat{\tau}(z, t) & = \tfrac{1}{\vel(t + \hat{\tau}(z,t))}, &
    \hat{\tau}(0, t) & = 0. \label{eq:control:predictor:tau}
\end{align}
Analogously to \eqref{eq:charSys:delsys:int}, the quantities $\bmw\Pred(z,t) = \bmw(z, \hat{\varphi}(z,t))$ satisfy \eqref{eq:model:del} with
$\vel\Del(z,t)$ replaced by $\vel\Pred(z,t) = \vel(\hat{\varphi}(z,t))$, where the boundary values at $z = 0$ are the current inputs $\ctrlIn(t)$ and
$\bdistIn(t)$ and $\bmW \bmZeta\Pred(L,t) = \ctrlOut\Pred(t)$ must be predicted\footnote{
    Alternatively, the forward problem \eqref{eq:model:del} could be used for prediction, as described, e.g., in \cite{Fridman2014}.
    However, this requires knowledge of the process state over the entire spatial domain $z\in \spatDom$ at time $t$, which would necessitate the
    use of an additional observer.
}.
To this end, a predictor-corrector approach similar to the Smith predictor \cite{Smith1957} is proposed.

In parallel with the actual process, a predictor is employed, which operates as a simulator of the model on the forward characteristics.
In analogy to \eqref{eq:ff:nonlin:transport:sol}, the solution of the transport equations is written as
\begin{align*}
    \bar{\hat{\bmZeta}}(z) & = \cOdesFlow[$\hat{\bmf}$](z, \bar{\ctrlIn}, \bar{\bdistIn}, \chi_{[0,z]}\bar{\hat{\Eta}}, \chi_{[0,z]}\bar{\vel}),
\end{align*}
using the map $\cOdesFlow[$\hat{\bmf}$] \colon \spatDom \times \Re^{n} \times \Cont[$0$](\spatDom) \times \Cont[$1$](\spatDom)
\varrightarrow \Re^{n}$, where the boundary values $\bar{\ctrlIn}$ and $\bar{\bdistIn}$ determine $\bar{\hat{\bmZeta}}(0)$ via
\eqref{eq:model:inout}.
Substituting this solution into \eqref{eq:model:del:storage} yields the predictor
\begin{equation}
    \hspace{-0.5ex}\pderi[t] \hat{\Eta}\Pred(z,t) = \tilde{\stg}(z, \ctrlIn(t), \bdistIn(t), \hat{\Eta}\Pred(\cdot,t), \vel\Pred(\cdot,t), \vel(t))
    \label{eq:control:predictor}
\end{equation}
with the function $\tilde{\stg} \colon \spatDom \times \Re^{n} \times \Cont[$0$](\spatDom) \times \Cont[$1$](\spatDom) \times \Re \varrightarrow \Re$,
\begin{multline*}
    \tilde{\stg}(z, \bar{\ctrlIn}, \bar{\bdistIn}, \bar{\hat{\Eta}}, \bar{\vel}, \check{\vel}) \\
    = \hat{\stg}\big(\cOdesFlow[$\hat{\bmf}$](z, \bar{\ctrlIn}, \bar{\bdistIn}, \chi_{[0,z]}\bar{\hat{\Eta}}, \chi_{[0,z]}\bar{\vel}),
    \bar{\hat{\Eta}}(z), \bar{\vel}(z), \check{\vel}\big),
\end{multline*}
which is solved forward in time using the input $\ctrlIn(t)$ applied to the process.
This simulation continuously yields the predicted profile $\hat{\Eta}\Pred(\cdot, t)$ and thus the predicted output
\begin{align*}
    \hat{\ctrlOut}\Pred(t) & = \bmW \hat{\bmZeta}\Pred(L,t),
\end{align*}
i.e., the model-based estimate of $\ctrlOut(t + \hat{\tau}(L,t))$.

Since the predictor is only based on the model, $\hat{\ctrlOut}\Pred(t)$ deviates from the actual future output due to unknown disturbances and model
errors.
For the correction step, the prediction $\hat{\ctrlOut}(t)$ of the current output is required, i.e., the value $\hat{\ctrlOut}\Pred(\check{t})$
computed at the earlier time instant $\check{t}$ with $\check{t} + \hat{\tau}(L,\check{t}) = t$.
The deviation
\begin{align}
    \Updelta \ctrlOut(t) & = \ctrlOut[m](t) - \hat{\ctrlOut}(t)
    \label{eq:control:out}
\end{align}
between measurement and prediction is assumed to remain constant over the prediction horizon $[t, t + \hat{\tau}(L,t)]$.
This yields the corrected predicted output
\begin{align}
    \breve{\ctrlOut}\Pred(t) & = \hat{\ctrlOut}\Pred(t) + \Updelta \ctrlOut(t) = \hat{\ctrlOut}\Pred(t) - \hat{\ctrlOut}(t) + \ctrlOut[m](t),
    \label{eq:control:corrPred}
\end{align}
which is made available to the controller.

\subsection{Feedback law}

The feedback is designed for the tracking error
\begin{align}
    \invbreve{\ctrlOut}(t) & = \breve{\ctrlOut}\Pred(t) - \ctrlOut[r]\Pred(t), &
    \ctrlOut[r]\Pred(t) & = \ctrlOut[r](t + \hat{\tau}(L,t)),
    \label{eq:control:error}
\end{align}
which is defined at the future time instant at which the current input becomes effective.
In the following, the \textit{uncoupled} case is considered first.
On the forward characteristics, the time derivative of the predicted output can be determined in the same way as \eqref{eq:ff:nonlin:dty:uncoupled},
\begin{multline}
    \dot{\hat{\ctrlOut}}\Pred(t) = \psi\big(\ctrlIn(t), \bdistIn(t), \dtbdistIn(t), \hat{\Eta}\Pred(\cdot,t), \\
    \vel\Pred(\cdot,t), \dot{\vel}\Pred(\cdot,t), \vel(t)\big),
    \label{eq:control:virtual}
\end{multline}
where the function $\psi$ results from differentiating $\hat{\ctrlOut}\Pred(t)$ w.r.t. time analogously to \eqref{eq:ff:nonlin:dty:uncoupled}
and additionally contains the terms due to the time-variant velocity, involving $\dot{\vel}\Pred(z,t) = \dot{\vel}(\hat{\varphi}(z,t))$.
Due to condition \eqref{eq:ff:nonlin:inputCondRel}, the derivative of $\psi$ w.r.t. $\ctrlIn(t)$ does not vanish.
Interpreting the right-hand side of \eqref{eq:control:virtual} as the virtual input $w(t)$ yields the integrator $\dot{\hat{\ctrlOut}}\Pred(t) =
w(t)$, and, by the implicit function theorem, the input
\begin{multline}
    \ctrlIn(t) = h_{\ttext{u}}\big(w(t), \bdistIn(t), \dtbdistIn(t), \hat{\Eta}\Pred(\cdot,t), \\
    \vel\Pred(\cdot,t), \dot{\vel}\Pred(\cdot,t), \vel(t)\big)
    \label{eq:control:inverse}
\end{multline}
that realises a given $w(t)$ locally\footnote{
    Here, $h_{\ttext{u}}$ generalises \eqref{eq:ff:nonlin:uncoupled:input} to a time-variant velocity.
}.
With the integrated tracking error $e(t) = \int_{\tS}^{t} \invbreve{\ctrlOut}(\vartheta) \dd \vartheta$, the virtual input is chosen as
\begin{equation}
    w(t) = \dtCtrlOut[r]\Pred(t) + \bar{w}(t),\quad
    \bar{w}(t) = - \Kp\, \invbreve{\ctrlOut}(t) - \Ki\, e(t)
    \label{eq:control:w}
\end{equation}
including a proportional part as well as an integral part for steady-state accuracy.
Assuming a constant deviation $\Updelta \ctrlOut$ in \eqref{eq:control:corrPred}, $\dot{\breve{\ctrlOut}}\Pred(t) = \dot{\hat{\ctrlOut}}\Pred(t) =
w(t)$ holds, which leads to the linear tracking-error dynamics
\begin{align*}
    \ddot{e}(t) + \Kp\, \dot{e}(t) + \Ki\, e(t) & = 0,
\end{align*}
whose characteristic polynomial is Hurwitz for $\Kp, \Ki \in \Re^{+}$.

In the \textit{coupled} case, the relative degree is $r = 0$ and the predicted output depends directly on the input.
Evaluating \eqref{eq:ff:nonlin:output} on the forward characteristics gives
\begin{align*}
    \hat{\ctrlOut}\Pred(t) & = \bmW \cOdesFlow[$\hat{\bmf}$]\big(L, \ctrlIn(t), \bdistIn(t), \hat{\Eta}\Pred(\cdot,t), \vel\Pred(\cdot,t)\big),
\end{align*}
which, due to \eqref{eq:ff:nonlin:inputCond}, can be solved locally for the input, i.e.,
\begin{align}
    \ctrlIn(t) & = h_{\ttext{c}}\big(w(t), \bdistIn(t), \hat{\Eta}\Pred(\cdot,t), \vel\Pred(\cdot,t)\big).
    \label{eq:control:inverse:coupled}
\end{align}
In contrast to \eqref{eq:control:inverse}, the virtual input $w(t) = \hat{\ctrlOut}\Pred(t)$ now directly specifies the predicted output.
With
\begin{align}
    w(t) & = \ctrlOut[r]\Pred(t) + \bar{w}(t), &
    \bar{w}(t) & = - \Ki\, e(t)
    \label{eq:control:w:coupled}
\end{align}
and a constant deviation $\Updelta \ctrlOut$, the tracking error \eqref{eq:control:error} becomes $\invbreve{\ctrlOut}(t) = \Updelta \ctrlOut -
\Ki\, e(t)$, i.e., $\dot{e}(t) + \Ki\, e(t) = \Updelta \ctrlOut$.
Hence, $\invbreve{\ctrlOut}(t) \varrightarrow 0$ for $\Ki \in \Re^{+}$, and a pure integral action suffices.
The structure in \figref{fig:ctrl:overview} remains unchanged, with $\dtCtrlOut[r]\Pred(t)$ replaced by $\ctrlOut[r]\Pred(t)$ and
\eqref{eq:control:inverse} by \eqref{eq:control:inverse:coupled}.

For $\bar{w} \equiv 0$ and $\hat{\Eta}\Pred(\cdot, t) = \Eta[r]\Pred(\cdot, t)$, the inversions \eqref{eq:control:inverse} and
\eqref{eq:control:inverse:coupled} reproduce the feedforward laws \eqref{eq:ff:nonlin:uncoupled:input} and the one resulting from
\eqref{eq:ff:nonlin:coupled:bvp:sol}, respectively, evaluated on the forward instead of the backward characteristics.
Hence, in both cases, the controller only compensates for the deviations from the feedforward.
As in the feedforward design, it is essential that the predictor is stable, since its dynamics under the inversion \eqref{eq:control:inverse}
or \eqref{eq:control:inverse:coupled} correspond to the internal dynamics.

\section{Example}
\label{sec:example}
The proposed control concept is applied to an example system of the form \eqref{eq:model}.
It comprises two transport processes $\bmZeta(z,t) = (\Zeta[1](z,t), \Zeta[2](z,t))^{\trans}$ as well as the nonlinear reaction terms
\begin{align*}
    \stf[1](\bar{\bmZeta}, \bar{\Eta}) & = \alpha_1\, \bar{\Zeta}_1 \left(1 - \bar{\Eta}\right) + \big(\alpha_2 \, \bar{\Zeta}_2 + \alpha_3 \big) \bar{\Eta} \\
    \stf[2](\bar{\bmZeta}, \bar{\Eta}) & = \big(\beta_1\, \bar{\Zeta}_1 + \beta_2\, \bar{\Zeta}_2\big) \bar{\Eta} \\
    \stg(\bar{\bmZeta}, \bar{\Eta}) & = \gamma_1\, \bar{\Zeta}_1 (1 - \bar{\Eta}) - \big( \gamma_2 + \gamma_3 \bar{\Zeta}_2 \big) \bar{\Eta}
\end{align*}
with the parameters in \tabref{tab:example}.
\begin{table}[b]
    \caption{Parameter overview for the considered example.}
    \vspace{-3ex}
    \label{tab:example}
    \begin{center}
        \begin{tabular}{c|ccc|cc|ccc}
            $L$ & $\alpha_1$ & $\alpha_2$ & $\alpha_3$ & $\beta_1$ & $\beta_2$ & $\gamma_1$ & $\gamma_2$ & $\gamma_3$ \\
            \midrule
            \num{5} & \num{-0.5} & \num{-0.03} & \num{3} & \num{-0.08} & \num{-0.05} & \num{0.05} & \num{0.3} & $\num{0.005}$ \\
        \end{tabular}
    \end{center}
\end{table}
The input, the disturbance, and the output, i.e., the measurement, are given according to \eqref{eq:model:inout} by the vectors
\begin{align*}
    \bmVo & = \begin{pmatrix}
        1 & 0 \\
    \end{pmatrix}, &
    \bmV & = \begin{pmatrix}
        0 & 1 \\
    \end{pmatrix}, &
    \bmW & = \begin{pmatrix}
        0 & 1
    \end{pmatrix}.
\end{align*}
First, the coupling structure of the reaction terms of the transport subsystem is examined by evaluating the Jacobian matrix w.r.t. $\bar{\bmZeta}$,
\begin{align*}
    \pderi[\bar{\bmZeta}] \bmf(\bar{\bmZeta}, \bar{\Eta}) & = \begin{pmatrix}
        \alpha_1 \left(1 - \bar{\Eta}\right) & \alpha_2 \, \bar{\Eta} \\
        \beta_1\, \bar{\Eta} & \beta_2\, \bar{\Eta}
    \end{pmatrix}.
\end{align*}
The control input $\ctrlIn(t)$ acts directly on the output $\ctrlOut(t)$ only via $\beta_1$, since $\Zeta[2](\cdot, t)$ depends on $\Zeta[1](\cdot, t)$
through this parameter, whereas $\alpha_2$ describes the reverse coupling of $\Zeta[2](\cdot, t)$ to $\Zeta[1](\cdot, t)$.
Hence, $\beta_1 = 0$ is sufficient for the input to act on the output only via the storage variable.
Moreover, for $\alpha_2 = \beta_1 = 0$, the two transport equations are completely decoupled.
Accordingly, the \textit{coupled} and \textit{uncoupled} cases can be distinguished.

The process model \eqref{eq:model} and the characteristic system \eqref{eq:model:del} used for the predictor and the feedforward are discretised
separately, since they serve different purposes.
The process model \eqref{eq:model}, acting as the simulated plant, is spatially discretised directly, using the backward difference $\pderi[z]
\Zeta(z,t) = \frac{\Zeta[$i$](t) - \Zeta[$i-1$](t)}{\Updelta z}$ in the sense of the \gls{fdm}, with step size $\Updelta z = \nicefrac{L}{(N_z - 1)}$
for $N_z$ spatial discretisation points.
The resulting system of \glspl{ode} is then solved using a standard \gls{ode} solver.

For the predictor and the feedforward, however, determining the analytical solutions from \secref{sec:ff:nonlin:coupled} and
\secref{sec:ff:nonlin:uncoupled} is avoided by instead spatially discretising the system directly on the characteristics \eqref{eq:model:del}, again
using the backward difference.
The resulting system is interpreted as a \gls{dde} system and solved by integrating along the characteristics, which are described by
\begin{align*}
    \deri[\tau]{s}(s, t) & = \tfrac{1}{\vel(t - \tau(s,t))}, & \tau(0, t) & = 0, & s & = L - z,
\end{align*}
and, for the predictor, by \eqref{eq:control:predictor:tau}.
In both cases, the resulting \gls{dde} system, with its non-equidistant time steps, is integrated using an explicit forward Euler method.

For the process model, $N_z = \num{101}$ spatial discretisation points are used, while a coarser discretisation with $N_z = \num{21}$ points suffices
for both the feedforward and the predictor.

To test the closed loop with the structure shown in \figref{fig:ctrl:overview}, the input $\ctrlIn(t)$ according to \eqref{eq:control:inverse} and \eqref{eq:control:inverse:coupled}, respectively, is
subjected to a constant error $\delta \ctrlIn = \num{0.1}$, i.e., the process is driven by $\ctrlIn(t) + \delta \ctrlIn$, where $\delta \ctrlIn$ is
unknown to the inverse model and the predictor.
The implementations were carried out in Python.

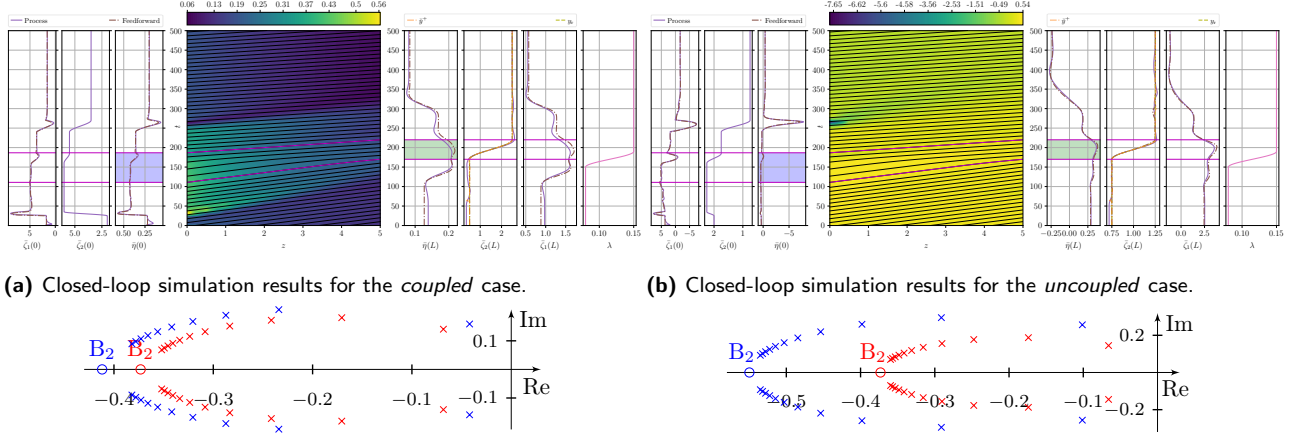
\begin{figure*}[t]
    \centering
    \begin{subfigure}[t]{0.5\textwidth}
        \includegraphics[width=\textwidth]{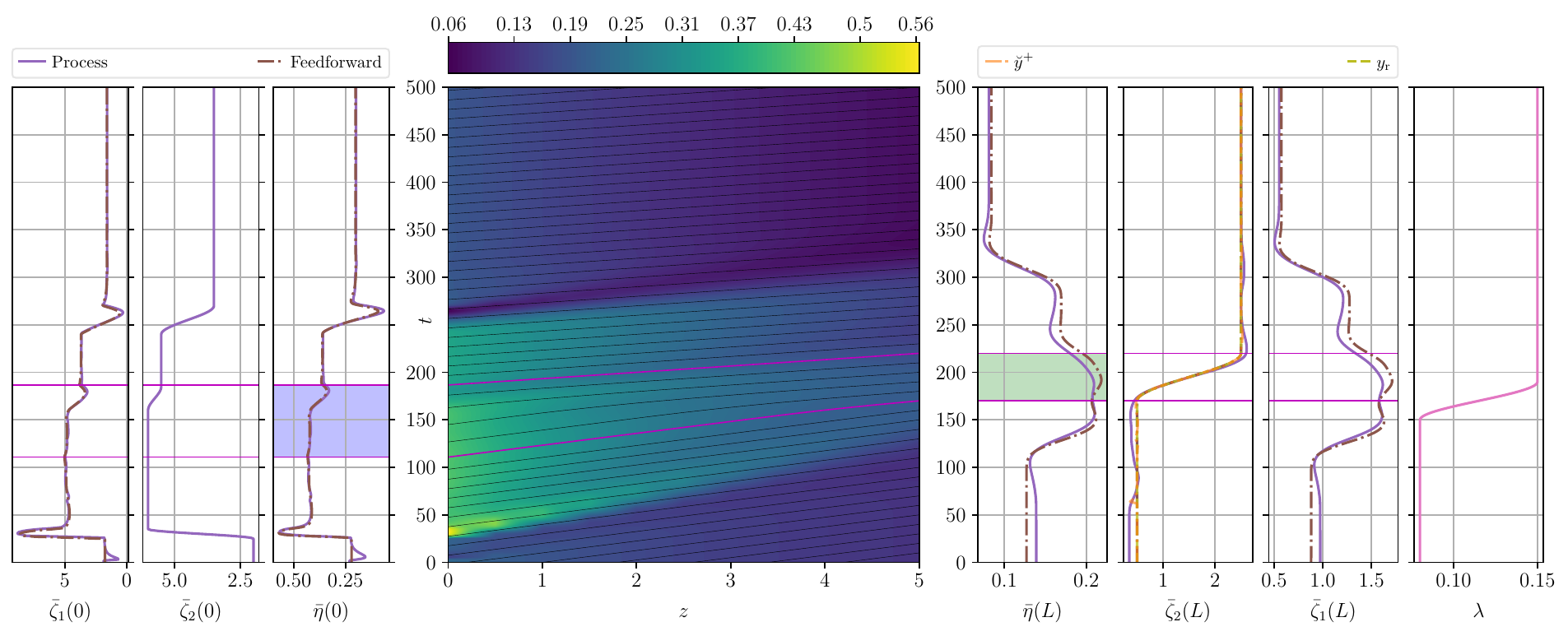}
        \caption{Closed-loop simulation results for the \textit{coupled} case.}
        \label{fig:example:coupled}
    \end{subfigure}%
    \begin{subfigure}[t]{0.5\textwidth}
        \includegraphics[width=\textwidth]{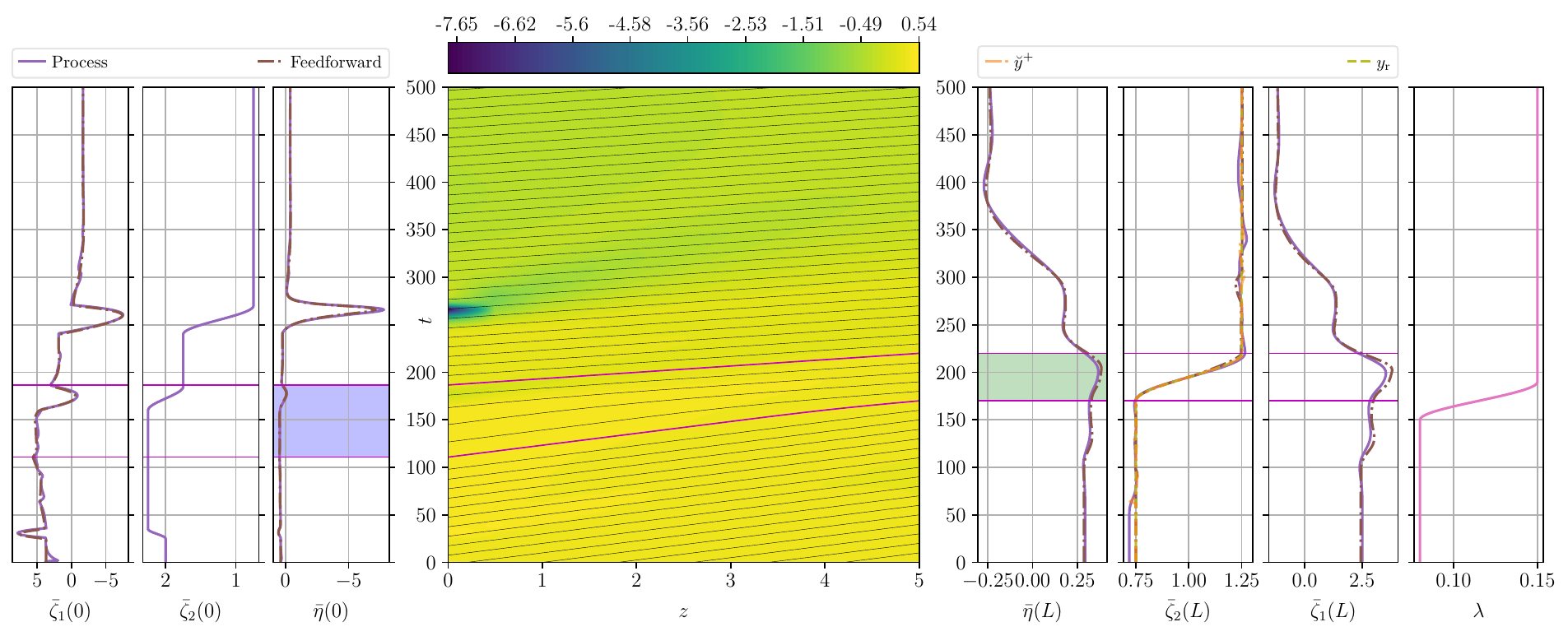}
        \caption{Closed-loop simulation results for the \textit{uncoupled} case.}
        \label{fig:example:uncoupled}
    \end{subfigure}
    \begin{subfigure}[t]{0.5\textwidth}
        \centering
        \begin{adjustbox}{width=0.75\textwidth}
            \begin{tikzpicture}[auto, >=latex']
    \pgfplotstableread[col sep=comma, header=false]{figures/example2ndNonLinCoupled.csv}\dataTable
    \begin{axis}[
        axis x line=middle,
        axis y line=middle,
        x axis line style={->},
        y axis line style={->},
        xlabel={Re},
        ylabel={Im},
        xmin=-0.43,
        xmax=0.03,
        ymin=-0.21,
        ymax=0.21,
        every tick/.style={
        black,
        semithick,
        },
        xticklabel style={
            /pgf/number format/fixed,
            yshift=-1ex,
            font=\footnotesize,
        },
        yticklabel style={
            /pgf/number format/fixed,
            /pgf/number format/precision=5,
            font=\footnotesize,
        },
        scaled y ticks=false,
        xtick={-0.4, -0.3, -0.2, -0.1},
        ytick={-0.1, 0.1},
        width=\textwidth,
        height=0.4\textwidth,
        xlabel style={at=(current axis.right of origin), anchor=north, xshift=-0.45ex},
        ylabel style={at=(current axis.above origin), anchor=west, yshift=-1ex},
    ]

        \def\coordsBlueX{}
        \def\coordsBlueO{}
        \def\coordsRedX{}
        \def\coordsRedO{}

        \pgfplotsforeachungrouped \i in {0,...,21}{
            \pgfplotstablegetelem{\i}{0}\of\dataTable
            \let\xVal=\pgfplotsretval
            \pgfplotstablegetelem{\i}{1}\of\dataTable
            \let\yVal=\pgfplotsretval
            \edef\coordsBlueX{\coordsBlueX (\xVal,\yVal) }
        }
        \pgfplotstablegetelem{22}{0}\of\dataTable
        \let\xVal=\pgfplotsretval
        \pgfplotstablegetelem{22}{1}\of\dataTable
        \let\yVal=\pgfplotsretval
        \edef\coordsBlueO{(\xVal,\yVal)}

        \pgfplotsforeachungrouped \i in {23,...,44}{
            \pgfplotstablegetelem{\i}{0}\of\dataTable
            \let\xVal=\pgfplotsretval
            \pgfplotstablegetelem{\i}{1}\of\dataTable
            \let\yVal=\pgfplotsretval
            \edef\coordsRedX{\coordsRedX (\xVal,\yVal) }
        }
        \pgfplotstablegetelem{45}{0}\of\dataTable
        \let\xVal=\pgfplotsretval
        \pgfplotstablegetelem{45}{1}\of\dataTable
        \let\yVal=\pgfplotsretval
        \edef\coordsRedO{(\xVal,\yVal)}

        \addplot[only marks, mark=x, blue] coordinates {\coordsBlueX};
        \addplot[only marks, mark=o, blue] coordinates {\coordsBlueO} node [yshift=2ex, xshift=0.0ex] {$\B[2]$};
        \addplot[only marks, mark=x, red]  coordinates {\coordsRedX};
        \addplot[only marks, mark=o, red]  coordinates {\coordsRedO} node [yshift=2ex, xshift=0.0ex] {$\B[2]$};

    \end{axis}

\end{tikzpicture}
        \end{adjustbox}
        \caption{Spectrum of the system operator for the stationary regimes at $\tsA$ in blue and $\tsB$ in red for the \textit{coupled} case.}
        \label{fig:example:coupled:eigenvals}
    \end{subfigure}%
    \begin{subfigure}[t]{0.5\textwidth}
        \centering
        \begin{adjustbox}{width=0.75\textwidth}
            \begin{tikzpicture}[auto, >=latex']
    \pgfplotstableread[col sep=comma, header=false]{figures/example2ndNonLinUnCoupled.csv}\dataTable
    \begin{axis}[
        axis x line=middle,
        axis y line=middle,
        x axis line style={->},
        y axis line style={->},
        xlabel={Re},
        ylabel={Im},
        xmin=-0.58,
        xmax=0.03,
        ymin=-0.32,
        ymax=0.32,
        every tick/.style={
        black,
        semithick,
        },
        xticklabel style={
            /pgf/number format/fixed,
            yshift=-1ex,
            font=\footnotesize,
        },
        yticklabel style={
            /pgf/number format/fixed,
            /pgf/number format/precision=5,
            font=\footnotesize,
        },
        scaled y ticks=false,
        xtick={-0.5, -0.4, -0.3, -0.2, -0.10},
        ytick={-0.2, 0.2},
        width=\textwidth,
        height=0.4\textwidth,
        xlabel style={at=(current axis.right of origin), anchor=north, xshift=-0.45ex},
        ylabel style={at=(current axis.above origin), anchor=west, yshift=-1ex},
    ]

        \def\coordsBlueX{}
        \def\coordsBlueO{}
        \def\coordsRedX{}
        \def\coordsRedO{}

        \pgfplotsforeachungrouped \i in {0,...,21}{
            \pgfplotstablegetelem{\i}{0}\of\dataTable
            \let\xVal=\pgfplotsretval
            \pgfplotstablegetelem{\i}{1}\of\dataTable
            \let\yVal=\pgfplotsretval
            \edef\coordsBlueX{\coordsBlueX (\xVal,\yVal) }
        }
        \pgfplotstablegetelem{22}{0}\of\dataTable
        \let\xVal=\pgfplotsretval
        \pgfplotstablegetelem{22}{1}\of\dataTable
        \let\yVal=\pgfplotsretval
        \edef\coordsBlueO{(\xVal,\yVal)}

        \pgfplotsforeachungrouped \i in {23,...,44}{
            \pgfplotstablegetelem{\i}{0}\of\dataTable
            \let\xVal=\pgfplotsretval
            \pgfplotstablegetelem{\i}{1}\of\dataTable
            \let\yVal=\pgfplotsretval
            \edef\coordsRedX{\coordsRedX (\xVal,\yVal) }
        }
        \pgfplotstablegetelem{45}{0}\of\dataTable
        \let\xVal=\pgfplotsretval
        \pgfplotstablegetelem{45}{1}\of\dataTable
        \let\yVal=\pgfplotsretval
        \edef\coordsRedO{(\xVal,\yVal)}

        \addplot[only marks, mark=x, blue] coordinates {\coordsBlueX};
        \addplot[only marks, mark=o, blue] coordinates {\coordsBlueO} node [yshift=2ex, xshift=-0.75ex] {$\B[2]$};
        \addplot[only marks, mark=x, red]  coordinates {\coordsRedX};
        \addplot[only marks, mark=o, red]  coordinates {\coordsRedO} node [yshift=2ex, xshift=-0.75ex] {$\B[2]$};

    \end{axis}

\end{tikzpicture}
        \end{adjustbox}
        \caption{Spectrum of the system operator for the stationary regimes at $\tsA$ in blue and $\tsB$ in red for the \textit{uncoupled} case.}
        \label{fig:example:uncoupled:eigenvals}
    \end{subfigure}%
    \caption{
        Simulation results of the closed loop for both variants for a velocity $\vel(t)$ and disturbance profile $\Zeta[2](0, t)$ as well as a
        transition of the reference variable $\ctrlOut[r](t)$ for the output $\ctrlOut(t) = \Zeta[2](L,t)$, shown by the magenta lines (\plotMagentaLine) in which the process
        variables at the input and output are compared with those of the feedforward, and the temporal and spatial evolution of the storage variable
        $\Eta(z,t)$ on the characteristics is presented.
    }
    \label{fig:example}
    \vspace{-4ex}
\end{figure*}

\subsection{Coupled case}

For the \textit{coupled} case $(\beta_1 \neq 0)$, a transition of the reference $\ctrlOut[r](t) = \Zeta[2](L, t)$ is planned from
$\ctrlOut[r](\tsA) = \num{0.5}$ to $\ctrlOut[r](\tsB) = \num{2.5}$ over the time interval $t \in [\tsA, \tsB] = [\num{170},
\num{220}]$.
The a priori known trajectories of the disturbance $\distIn(t) = \Zeta[2](0, t)$ and of the velocity $\vel(t)$ can be seen in
\figref{fig:example:coupled}.
To compute the corresponding feedforward law, the discretised model is used.

For this purpose, the \gls{bvp} \eqref{eq:ff:nonlin:coupled:bvp} must first be solved on the characteristics.
This is achieved by discretising \eqref{eq:ff:nonlin:coupled:bvp} by means of the backward difference, which yields the system of equations
\begin{align}
    \bm{F}(\bmEta\Del(t), \vel\Del(t)) \bmZeta\Del(t) & = \bm{b}(\bmEta\Del(t), \vel\Del(t), t)
    \label{eq:example:ff:nonlin:coupled:bvp:sol}
\end{align}
with $\bm{F} \in \Re^{2 N_z \times 2 N_z}$, $\bmZeta\Del(t) = ((\bmZeta[1]\Del(t))^{\trans}, (\bmZeta[2]\Del(t))^{\trans})^{\trans} \in \Re^{2 N_z}$,
$\bmZeta[$j$]\Del(t) = (\Zeta[$j,0$]\Del(t), \ldots, \Zeta[$j,N_z-1$]\Del(t))^{\trans}$, $j = 1, 2$, and
$\bm{b} = \big(\distIn\Del(t), \tfrac{\Updelta z\, \alpha_3}{\vel[1]\Del(t)} \Eta[1]\Del(t), \ldots, \tfrac{\Updelta z\,
\alpha_3}{\vel[$N_z-1$]\Del(t)} \Eta[$N_z-1$]\Del(t), 0, \ldots, 0, \allowbreak \ctrlOut[r]\Del(t)\big)^{\trans}\in \Re^{2 N_z}$,
where the entries depending on $\bmEta\Del(t)$ result from the inhomogeneous term $\alpha_3\, \bar{\Eta}$ in $\stf[1]$.
The vector $\bmEta\Del(t) = (\Eta[0]\Del(t), \ldots, \Eta[$N_z - 1$]\Del(t))^{\trans}$ appearing therein is obtained by integrating
$\dot{\bmEta}\Del(t) = \tfrac{\vel(t)}{\vel\Del(t)}  \Big(\gamma_1\, \bmZeta[1]\Del(t) \big(1 - \bmEta\Del(t)\big) - \big( \gamma_2 + \gamma_3
\bmZeta[2]\Del(t) \big) \bmEta\Del(t)\Big)$ over time, using $\bmZeta[1]\Del(t)\in\Re^{N_z}$ and $\bmZeta[2]\Del(t)\in\Re^{N_z}$ from
\eqref{eq:example:ff:nonlin:coupled:bvp:sol}.
Therein, $\vel\Del(t)$ contains the delayed velocities at the grid points, so that the factor $\nicefrac{\vel(t)}{\vel\Del(t)}$ from
\eqref{eq:model:del:storage} is taken into account, and all products, quotients, and sums involving vectors are to be understood element-wise.

In order to assess the stability of the internal dynamics, the system is linearised around the stationary regimes corresponding to the
reference $\ctrlOut[r](\bar{t})$, the disturbance $\distIn(\bar{t})$, and the velocity $\vel(\bar{t})$ at the two time instants
$\bar{t} \in \{\tsA, \tsB\}$.
Since the stationary profiles are spatially varying, this linearisation leads to spatially dependent coefficients, cf. \secref{sec:problem}.
As an approximation, cf. \secref{sec:ff:nonlin:coupled}, these are frozen at the stationary values at the outlet $z = L$, denoted by
$\square^{\star}$, for the evaluation of the spectrum.
As a result, a system of the form \eqref{eq:ff:lin} is obtained, with the constant coefficients
$\hbmA[1] = \frac{1}{\vel^{\star}}\begin{psmallmatrix}
    \alpha_1 (1 - \Eta^{\star}) & \alpha_2 \Eta^{\star} \\
    \beta_1 \Eta^{\star} & \beta_2 \Eta^{\star}
\end{psmallmatrix}$,
$\bmA[2] = \begin{psmallmatrix}
    \gamma_1 (1 - \Eta^{\star}) & -\gamma_3 \Eta^{\star}
\end{psmallmatrix}$,
$\hbmB[1] = \frac{1}{\vel^{\star}}\begin{psmallmatrix}
    -\alpha_1 \Zeta[1]^{\star} + \alpha_2 \Zeta[2]^{\star} + \alpha_3 \\
    \beta_1 \Zeta[1]^{\star} + \beta_2 \Zeta[2]^{\star}
\end{psmallmatrix}$ and
$\B[2] = -\gamma_1\, \Zeta[1]^{\star} - \gamma_2 - \gamma_3 \Zeta[2]^{\star}$,
together with the stationary values $\Eta^{\star}$, $\Zeta[1]^{\star}$, $\Zeta[2]^{\star}$ at $z = L$ and the velocity $\vel^{\star} = \vel(\bar{t})$.
Based on this linear representation, the spectrum of the system operator can now be evaluated as described in \secref{sec:ff:lin:coupled}.
It corresponds to \eqref{eq:ff:lin:coupled:spec} and is shown in \figref{fig:example:coupled:eigenvals}, where the first \num{11} conjugate eigenvalue
pairs, together with $\B[2]$, are depicted in blue for $\tsA$ and in red for $\tsB$, respectively.
As can be seen, the computed eigenvalues lie in the left half-plane and approach the point $\B[2] < 0$.
This indicates that the internal dynamics are locally exponentially stable at the two stationary regimes considered.

Since the control input $\ctrlIn(t)$ is subjected to the constant error $\delta \ctrlIn$, as mentioned previously, the virtual input
\eqref{eq:control:w:coupled} with the integral gain $\Ki = \num{0.2}$ is employed to compensate for it.
Since $\delta \ctrlIn$ acts from the outset, the process starts in the stationary regime belonging to $\ctrlIn(0) + \delta \ctrlIn$, so that its
output initially deviates from the reference by about \num{0.15}.
This initial error must likewise be compensated by the controller.

The results of this scenario are shown in \figref{fig:example:coupled}.
In addition to the closed-loop process variables, the figure also shows the profiles of $\Zeta[1](\cdot,t)$ and $\Zeta[2](\cdot, t)$ at the input and
output, as computed by the feedforward.
The influence of the transport delay can be clearly observed at the beginning of the simulation, since the output does not respond to the
controller until after the delay time of $\tau(L,\tS) = \hat{\tau}(L,\tS) = \nicefrac{L}{\vel(\tS)} = \num{62.5}$, after
which the initial error is compensated within a period of about \num{15} owing to the integral action.
When the change of the disturbance at the inlet between $t = \num{25}$ and $t = \num{35}$ reaches the outlet, a deviation of up to \num{0.13}
occurs again.
It results from the different discretisations of the process model and the predictor, whose deviation \eqref{eq:control:out} thereby changes,
whereas it is assumed to be constant in the correction, so that this change is only compensated after the delay.
Such peaks are therefore limited by the delay and hardly depend on the gain.
A larger $\Ki$ merely shortens the subsequent compensation.
Nevertheless, the controller accounts for the constant disturbance and the a priori known disturbance and velocity trajectories, such that the
reference is reached with a minimal overshoot at the end of the transition, and the tracking error is practically zero from about $t = \num{370}$ on,
even though only a coarse discretisation was chosen for the predictor.

\subsection{Uncoupled case}

For the \textit{uncoupled} case, $\alpha_2 = \beta_1 = 0$ is chosen, i.e., the transport equations are completely decoupled.
In this scenario, a transition from $\ctrlOut[r](\tsA) = \num{0.75}$ to $\ctrlOut[r](\tsB) = \num{1.25}$ is planned over the time interval $t \in
[\tsA, \tsB] = [\num{170}, \num{220}]$.
\figref{fig:example:uncoupled} directly shows the a priori known trajectories of the disturbance $\distIn(t)$ and the velocity $\vel(t)$ that occur
over this interval.

To compute the feedforward law \eqref{eq:ff:nonlin:uncoupled:input}, the transport equations are first considered on the characteristics and
discretised by means of the backward difference.
This yields
\begin{subequations}\label{eq:example:uncoupled:zeta}
\begin{align}
    \bmZeta[1]\Del(t) & = \bm{F}_{1,1}(\bmEta\Del(t), \vel\Del(t)) \notag \\
                      & \quad + \bm{F}_{1,2}(\bmEta\Del(t), \vel\Del(t))\ctrlIn\Del(t) \\
    \bmZeta[2]\Del(t) & = \bm{F}_2(\bmEta\Del(t), \vel\Del(t)) \distIn\Del(t)
\end{align}
\end{subequations}
with $\bmZeta[1]\Del(t)=(\Zeta[$1,1$]\Del(t), \ldots, \Zeta[$1,N_z-1$]\Del(t))^{\trans}$, $\bmZeta[2]\Del(t)=(\Zeta[$2,1$]\Del(t), \ldots,
\Zeta[$2,N_z-1$]\Del(t))^{\trans}$ and $\bmEta\Del(t)=(\Eta[1]\Del(t), \ldots, \Eta[$N_z - 1$]\Del(t))^{\trans}$, as well as $\bm{F}_{1,1},
\bm{F}_{1,2}, \bm{F}_{2} \in \Re^{N_z-1}$.
In contrast to the \textit{coupled} case, the boundary values $\Zeta[$1,0$]\Del(t) = \ctrlIn\Del(t)$ and $\Zeta[$2,0$]\Del(t) = \distIn\Del(t)$ are
not part of the state.
Next, the output equation $\ctrlOut[r]\Del(t) = \bm{e}^{\trans} \bmZeta[2]\Del(t)$, with the unit vector $\bm{e} = (0, \ldots, 0, 1)^{\trans}
\in \Re^{N_z - 1}$ selecting the outlet $z = L$, is differentiated once w.r.t. time.
Substituting
\begin{equation}
    \dot{\bmEta}\Del(t) \!=\! \tfrac{\vel(t)}{\vel\Del(t)} \big(\gamma_1 \bmZeta[1]\Del(t) \big(1 \!-\! \bmEta\Del(t)\big) \!-\! \big( \gamma_2
    \!+\! \gamma_3 \bmZeta[2]\Del(t) \big) \bmEta\Del(t)\big)
    \label{eq:example:uncoupled:eta}
\end{equation}
and \eqref{eq:example:uncoupled:zeta} into this expression then yields an equation that can be solved explicitly for the control input,
\begin{equation}
    \hspace{-0.5ex}\ctrlIn\Del(t) = \tfrac{\dtCtrlOut[r]\Del(t) - Q_1(\bmEta\Del(t), \lam\Del(t), \dot{\lam}\Del(t), \lam(t), \distIn\Del(t),
    \dot{\distIn}\Del(t))}{Q_2(\bmEta\Del(t), \lam\Del(t), \lam(t), \distIn\Del(t))}.
    \label{eq:example:uncoupled:ctrlIn}
\end{equation}
This expression involves $\bmEta\Del(t)$, as well as the a priori known disturbance $\distIn\Del(t)$ and the velocity $\vel\Del(t)$, together with
their time derivatives, and the current velocity $\vel(t)$.
As in the \textit{coupled} case, $\bmEta\Del(t)$ is determined by integrating \eqref{eq:example:uncoupled:eta} forward in time, using
$\bmZeta[1]\Del(t)$ and $\bmZeta[2]\Del(t)$ from \eqref{eq:example:uncoupled:zeta} together with $\ctrlIn\Del(t)$ from
\eqref{eq:example:uncoupled:ctrlIn}.

To assess stability, the same approach as in the \textit{coupled} case is followed, considering the stationary regimes at $\tsA$ and
$\tsB$.
In this case, however, the coefficients $\hbmA[1]$ and $\hbmB[1]$ from the \textit{coupled} case change to
$\hbmA[1] = \tfrac{1}{\vel^{\star}}\begin{psmallmatrix}
    \alpha_1 (1 - \Eta^{\star}) & 0 \\
    0 & \beta_2 \Eta^{\star}
\end{psmallmatrix}$ and
$\hbmB[1] = \tfrac{1}{\vel^{\star}}\begin{psmallmatrix}
    -\alpha_1 \Zeta[1]^{\star} + \alpha_3 \\
    \beta_2 \Zeta[2]^{\star}
\end{psmallmatrix}$.
The spectrum of the system operator is then computed as described in \secref{sec:ff:lin:uncoupled}, again corresponding to
\eqref{eq:ff:lin:coupled:spec}.
\figref{fig:example:uncoupled:eigenvals} depicts the eigenvalues at $\tsA$ in blue and those at $\tsB$ in red.
For both stationary regimes, the first \num{11} conjugate eigenvalue pairs lie in the left half-plane and approach the point $\B[2] < 0$, which likewise
indicates local exponential stability of the internal dynamics.

As in the \textit{coupled} case, the control input $\ctrlIn(t)$ is affected by a constant disturbance $\delta \ctrlIn$.
This is compensated by the virtual input \eqref{eq:control:w} with the gains $\Kp = \num{0.2}$ and $\Ki = \num{0.01}$, which
place a double pole of the tracking-error dynamics at $\num{-0.1}$.
Here too, the process starts in the stationary regime belonging to $\ctrlIn(0) + \delta \ctrlIn$, which results in an initial error of about
\num{0.03} that must also be compensated.

The resulting simulation is shown in \figref{fig:example:uncoupled}, where, alongside the closed-loop process variables, the feedforward-computed
profiles of $\Zeta[1](\cdot, t)$ and $\Zeta[2](\cdot, t)$ at the inlet and outlet are also included.
The results demonstrate that the initial error is compensated within a period of about \num{10}, once the delay time of $\tau(L,\tS) = \hat{\tau}(L,\tS) =
\num{62.5}$ has passed, after which the reference trajectory is tracked very closely.
The outlet value $\Zeta[2](L,t)$ computed by the feedforward also follows the reference closely after the velocity change.
Only after the change of the disturbance at the inlet between $t = \num{240}$ and $t = \num{270}$ does a small deviation occur at the outlet.
It results from the comparatively coarse discretisation, both spatially, with $N_z = \num{21}$ grid points, and temporally, with a step size of
\num{1}, since the feedforward integrates the time derivative of the output at $z=L$ and does not correct deviations of the output itself.
The controller, by contrast, corrects such deviations through the feedback, so that its tracking error vanishes towards the end of the simulation.

\section{Conclusion}
\label{sec:conclusion}

This article presents an inversion-based feedforward control and a predictor-based tracking controller for $n$ convection-reaction equations
that are coupled in a distributed manner with one reaction equation, first for a linear representation and then for nonlinear reaction terms.
Formulated on the characteristics, the coupling structure and the choice of inputs and outputs determine the relative degree, which is zero in
the \textit{coupled} and one in the \textit{uncoupled} case.
In the linear case, the resulting internal dynamics can be written as a well-posed abstract evolutionary \gls{ode}, whose stability can be
assessed by means of the spectrum of the bounded system operator.
Based on the inversion, the tracking controller, which compensates for uncertainties and unknown disturbances, is obtained by input-output
linearisation, and the transport delays are taken into account by a predictor.
The simulation results indicate that, owing to the formulation on the characteristics, a considerably coarser spatial discretisation suffices
for the feedforward and the predictor than for the process model.

The presented approach for a single reaction equation will be extended to the multivariable case with several nonlinear reaction equations.
Furthermore, the discretisation scheme used in \secref{sec:example} will be improved by introducing a separation into a stationary and a dynamic
component.
A similar approach was described in \cite{Bachler2017}.
In particular, the described representation as a nonlinear evolutionary \gls{ode} is also intended for use in developing a nonlinear
model predictive control scheme.

\vspace{-5ex}

\printbibliography

\vspace{-6ex}

\contributors{

\contributor{Dr. techn. Jens Wurm}{UMIT TIROL - Private University For Health Sciences and Health Technology, Institute of Automation and Control
Engineering, Eduard-Wallnöfer-Zentrum 1, \mbox{6060 Hall in Tirol,}
Austria}{jens.wurm@umit-tirol.at}{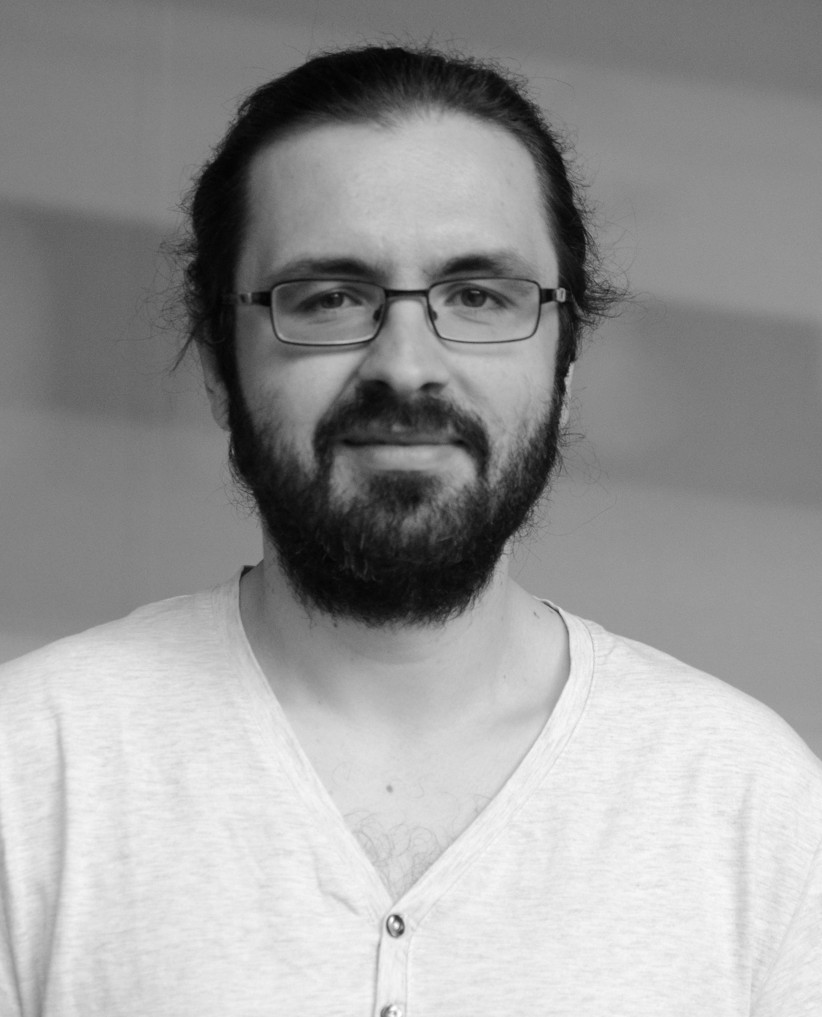}{%
    Jens Wurm received the Dr. techn. degree in technical sciences from UMIT TIROL in 2025.
    He is currently a Postdoctoral Researcher at the IACE, UMIT TIROL.
    His research interests include modelling, control and observer design for nonlinear infinite-dimensional systems.
}

\contributor{Univ-Prof. Dr.-Ing. Frank Woittennek}{UMIT TIROL - Private University For Health Sciences and Health Technology, Institute of Automation and Control
Engineering, Eduard-Wallnöfer-Zentrum 1, \mbox{6060 Hall in Tirol,}
Austria}{frank.woittennek@umit-tirol.at}{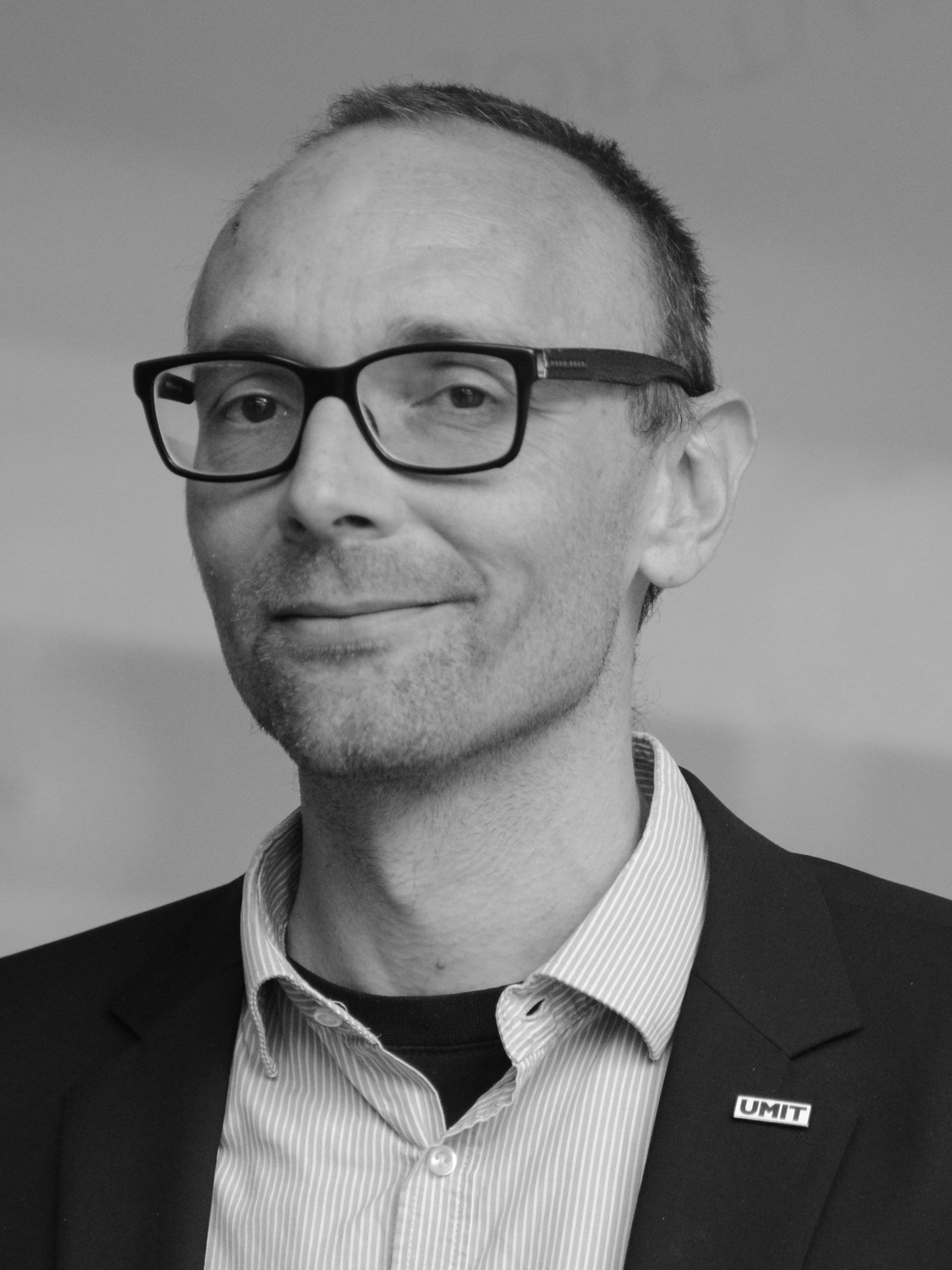}{%
    Frank Woittennek is head of the Institute of Automation and Control Engineering at UMIT TIROL in Hall in Tirol.
    His main research areas include control and observer design for linear and nonlinear finite- and infinite-dimensional systems, modelling, control
    and simulation, and state reconstruction for mechatronic, energy and process engineering systems.
}

}

\end{document}